\documentclass[a4paper,fleqn,usenatbib]{mnras}

\usepackage{newtxtext,newtxmath}

\usepackage[T1]{fontenc}
\usepackage{ae,aecompl}

\usepackage{graphicx}	
\usepackage{amsmath}	
\usepackage{amssymb}	

\newcommand{\Msol}{{\,\rm M}_\odot} 
\newcommand{\Zsun}{{\,\rm Z}_\odot}
\newcommand{\Mpc} {{\,\rm Mpc}}
\newcommand{\kpc} {{\,\rm kpc}} 

\newcommand{\pc} {{\,\rm pc}} 
\newcommand{\erg}{{\,\rm erg}}
\newcommand{\kms}{{\,\rm {km\,s^{-1}} }}

\newcommand{\cc}{{\,\rm {cm^{3}} }}

\title[What are damped HI absorbers?]{The nature of damped HI absorbers probed by cosmological simulations: satellite accretion and outflows}

\author[N. H. P. Rhodin et al.]{
N. H. P. Rhodin,$^{1}$\thanks{E-mail: henrikrhodin@dark-cosmology.dk}
O. Agertz,$^{2}$
L. Christensen,$^{1}$
F. Renaud,$^{2}$ and J. P. U. Fynbo,$^{3,4}$
\\
$^{1}$DARK, Niels Bohr Institute, University of Copenhagen, Lyngbyvej 2, 2100 Copenhagen \O, Denmark\\
$^{2}$Lund Observatory, Department of Astronomy and Theoretical Physics, Box 43, SE-221 00 Lund, Sweden\\
$^{3}$Cosmic Dawn Center (DAWN)\\ 
$^{4}$Niels Bohr Institute, University of Copenhagen, Lyngbyvej 2, 2100, Copenhagen \O, Denmark
}

\date{Accepted XXX. Received YYY; in original form ZZZ}

\pubyear{2019}

\begin{document}
\label{firstpage}
\pagerange{\pageref{firstpage}--\pageref{lastpage}}
\maketitle

\begin{abstract}
We use state-of-the-art cosmological zoom simulations to explore the distribution of neutral gas in and around galaxies that gives rise to high column density \ion{H}{i} \mbox{Ly-$\alpha$} absorption (formally, sub-DLAs and DLAs) in the spectra of background quasars. Previous cosmological hydrodynamic simulations under-predict the mean projected separations $(b)$ of these absorbers relative to the host, and invoke selection effects to bridge the gap with observations. On the other hand, single lines of sight (LOS) in absorption cannot uniquely constrain the galactic origin. Our simulations match all observational data, with DLA and sub-DLA LOS existing over the entire probed parameter space ($-4\lesssim $[M/H]$\lesssim 0.5$, $b<50$ kpc) at all redshifts ($z\sim 0.4 - 3.0$). We demonstrate how the existence of DLA LOS at $b\gtrsim 20-30$ kpc from a massive host galaxy require high numerical resolution, and that these LOS are associated with dwarf satellites in the main halo, stripped metal-rich gas and outflows. Separating the galaxy into interstellar (``\ion{H}{i} disc") and circumgalactic (``halo") components, we find that both components significantly contribute to damped \ion{H}{i} absorption LOS. Above the sub-DLA (DLA) limits, the disc and halo contribute with $\sim 60 (80)$ and $\sim 40 (20)$ per cent, respectively. Our simulations confirm analytical model-predictions of the DLA-distribution at $z\lesssim 1$. At high redshift ($z\sim 2-3$) sub-DLA and DLAs occupy similar spatial scales, but on average separate by a factor of two by $z\sim 0.5$. On whether sub-DLA and DLA LOS sample different stellar-mass galaxies, such a correlation can be driven by a differential covering-fraction of sub-DLA to DLA LOS with stellar mass. This preferentially selects sub-DLA LOS in more massive galaxies in the low-$z$ universe.
\end{abstract}

\begin{keywords}
galaxies: formation -- galaxies: evolution -- galaxies: haloes -- galaxies: intergalactic medium -- quasars: absorption lines
\end{keywords}



\section{Introduction}
\label{sec:introduction}

The chance alignment of neutral gas clouds intervening the line of sight (LOS) towards background quasi-stellar objects (quasars) imprints characteristic absorption features in the quasar power-law continua. In neutral hydrogen (\textsc{Hi}), the most \textsc{Hi}-rich absorbers are the Damped Lyman-$\alpha$ Absorbers (DLAs; \mbox{$\log _{10} [\mathrm{N}_\mathrm{\textsc{Hi}}~(\mathrm{cm}^{-2})] \geq 20.3$}, \cite{Wolfe1986}) and the sub-DLAs (\mbox{$19.0 \leq \log _{10} [\mathrm{N}_\mathrm{\textsc{Hi}}~(\mathrm{cm}^{-2})] < 20.3$}, e.g., \cite{Peroux2003,Zafar2013}). Unless otherwise specified, we refer to sub-DLAs and DLAs uniformly as damped \ion{H}{i} absorbers. Both classes of absorbers are always accompanied by low-ionization metal line complexes \citep{Prochaska2003,Noterdaeme2012b,Rafelski2014} which suggests an association with a gaseous medium affected by chemical enrichment. Yet, the origin and nature of damped \ion{H}{i} absorbers and their relation to galactic environments remains highly debated.

Absorption velocities and asymmetric line-profiles with leading edges \citep{Prochaska1997}, as well as models of disc formation \citep{Mo1998}, support the idea that DLAs preferentially probe rapidly rotating galaxy discs \citep{Wolfe1986}. However, using cosmological hydrodynamic simulations, \citet{Haehnelt1998} demonstrated that the absorption kinematics could equally well be explained by irregular proto-galactic clumps in dark matter halos subject to a combination of rotation, random motions, gas flows, and mergers. A picture is emerging in which damped \ion{H}{i} absorbers trace neutral gas on scales of tens of kpc in galaxies \citep{Moller1998,Christensen2007,Pontzen2008,Monier2009,Fynbo2010,Fynbo2011,Meiring2011,Rao2011,Krogager2012,Peroux2012,Fynbo2013,Krogager2013,Rahmati2014,Rahmani2016,Rhodin2018}. 

A number of analytical models have been successful in reproducing observed bulk properties of DLAs \citep[][from hereon the ``F08+K17 model"]{Fynbo2008, Krogager2017} \citep[also][in prep.]{Freudling2019}. By construction, these models depend on mean scaling relations of galaxies, and cannot capture process related to galactic sub-structure or gas structures in the intergalactic medium (IGM). Large-scale cosmological hydrodynamic simulations have the potential to mitigate these shortcomings. However, such simulations struggle with reproducing the observed absorption kinematics \citep{Bird2014, Bird2015}; lack resolution to capture cold gas in halos \citep{McCourt2018, Hummels2018}; and require very strong stellar feedback in order to produce cold gas at large impact parameters, destroying the star forming gas discs in the process \citep[][]{Liang2015}. Furthermore, work using cosmological simulations have called into question the detection of high impact parameter $(b~\mathrm{[\kpc ]}$) damped \ion{H}{i} absorbers, suggesting instead that observations suffer from a selection bias towards the most luminous (but unrelated) galaxy in the projected sky vicinity \citep{Rahmati2014}. We are thus left wondering what LOS through galactic environments that give rise to damped \ion{H}{i} absorbers.

In this work, we employ high resolution cosmological zoom simulations to build probability-functions of the distribution of damped \ion{H}{i} absorption in and around galaxies. These probability-functions are matched to the most recent compilation of spectroscopically confirmed absorber-galaxy pairs \citep[][from hereon, the ``MC19" compilation]{Moller2019, Christensen2019} and the analytical F08+K17 model, and are used to explore the physical origin of the damped \ion{H}{i} absorption. Throughout the analysis, we confine the comparison to a circular beam of $50~\mathrm{kpc}$ radius, centered on the galaxy. This value is observationally motivated, and selected to enclose the current compilation of detections. The paper is organized as follows: Section \ref{sec:simulations} describes our simulations, with particular emphasis given to the feedback recipes and resolution, Section \ref{sec:results} presents our results, and Section \ref{sec:conclusion} summarizes our conclusions. 

\section{Simulations}
\label{sec:simulations}

\subsection{Simulation setup}
\label{sec:setup}
We carry out a cosmological hydrodynamic+$N$-body zoom-in simulation of a Milky Way mass galaxy using the adaptive mesh refinement (AMR) code {\small RAMSES} \citep{teyssier02}, assuming a flat $\Lambda$-cold dark matter cosmology with \mbox{$H_0 = 70.2 \kms \Mpc^{-1}$}, $\Omega_{\rm m} = 0.272$, $\Omega_\Lambda = 0.728$, and \mbox{$\Omega_{\rm b}= 0.045$}. From a dark matter only simulation, with a simulation cube of size $L_{\rm box}=85 \Mpc$ at $z=0$, a halo of $R_{200,{\rm m}} = 334 \kpc$ (radius within which the mass density is 200 times the mean matter density) and $M_{200,{\rm m}} = 1.3\times 10^{12} \Msol$ was selected for re-simulation at high resolution. Particles within $3 R_{200,{\rm m}}$ at $z=0$ were traced back to $z=100$, and the Lagrangian region they defined was regenerated at high resolution, still embedded within the full lower-resolution volume, using the {\small MUSIC} code \citep{music2011}. The simulation was then run to $z=0$, with outputs every $\Delta a=0.01$. These are the same initial conditions as the ``m12i'' halo from \citet{Hopkins2014} and \citet{Wetzel2016}, drawn from the volume used in the AGORA galaxy formation comparison project \citep{agora, agora2}. 

The dark matter particle mass in the high resolution region is $m_{\rm dm}=3.5\times 10^4\Msol$, with a gas mass resolution of $7070\Msol$. Star formation is sampled using $10^4\Msol$ particles, with stellar evolution reducing this by up to $50\%$, see below. The adaptive mesh is allowed to refine if a cell contains more than eight dark matter particles. This allows the local force softening to closely match the local mean inter-particle separation, which suppresses discreteness effects \citep[e.g.,][]{Romeo08}. A similar criterion is employed for the baryonic component, where the maximum refinement level is set to allow for a mean constant physical resolution of $\sim 20 \pc$ in the dense interstellar medium. As such, this simulation is a significant improvement over previous cosmological zoom simulation used to study cold gas in and around galaxies \citep[e.g.][]{Liang2015}.

The adopted star formation and feedback physics is presented in \citet{Agertz2013} and \citet{AgertzKravtsov2015, agertzkravtsov2016}. Briefly, star formation is treated as a Poisson process occurring on a cell-by-cell basis according to the star formation law, 
\begin{equation}
	\dot{\rho}_{\star}= \epsilon_{\rm ff}\frac{ \rho_{\rm g}}{t_{\rm ff}},
	\label{eq:schmidtH2}	
\end{equation}
where $\dot{\rho}_{\star}$ is the star formation rate density, $\rho_{\rm g}$ the gas density, $t_{\rm ff}=\sqrt{3\pi/32G\rho_{\rm g}}$ is the local free-fall time and $\epsilon_{\rm ff}$ is the local star formation efficiency per free-fall time of gas in the cell. The efficiency is computed following the relation from \citet{Padoan2012}, derived from simulations of star formation in magnetized supersonic turbulence\footnote{$\epsilon_{\rm ff}=0.5\exp(-1.6 t_{\rm ff}/t_{\rm dyn})$, where the dynamical time is $t_{\rm dyn}=L/2\sigma$, and $\sigma$ is the local velocity dispersion compute using neighbouring gas cells over a region of size $L=$ 3 grid cells per spatial dimension.}. 

Each formed star particle is treated as a single-age stellar population with a \citet{chabrier03} initial mass function. We account for injection of energy, momentum, mass and heavy elements over time from Type II and Type Ia supernovae (SNe), stellar winds and radiation pressure (allowing for both single scattering and multiple scattering events on dust) on the surrounding gas. Each mechanism depends on the stellar age, mass and gas/stellar metallicity \citep[through the metallicity dependent age-mass relation of][]{Raiteri1996}, calibrated on the stellar evolution code {\small STARBURST99} \citep{Leitherer1999}. 

Furthermore, to accurately account for SN feedback we adopt the SN momentum injection model recently suggested by \citet[see also \citealt{Martizzi2015}]{KimOstriker2015}. A SN explosion is considered resolved when its cooling radius\footnote{The cooling radius in gas of density $n$ and metallicity $Z$ scales as $r\approx 30 (n/1\cc)^{-0.43} (Z/\Zsun + 0.01)^{-0.18} \pc$ for a supernova explosion with energy $E_{\rm SN}=10^{51}$ erg \citep[e.g.][]{Cioffi1988, Thornton1998}.} is captured by at least 6 grid cells. In this case the explosion is initialized in the energy conserving phase by injecting the relevant energy ($10^{51} \erg$ per SN) into the nearest grid cell. If this criterion is not fulfilled, the SN is initialized in its momentum conserving phase, i.e. the total momentum generated during the energy conserving Sedov-Taylor phase is injected into the cells surrounding a star particle. It can be shown \citep[e.g.][]{Blondin1998, KimOstriker2015} that at this time, the momentum of the expanding shell is approximately $4\times 10^5 (E_{\rm SN}/10^{51}\erg)^{16/17} (n/1~{\rm cm}^{-3})^{-2/17} {\Msol \kms}$.

We track iron (Fe) and oxygen (O) abundances separately, and advect them as passive scalars. When computing the gas cooling rate, which is a function of total metallicity, we construct a total metal mass as
\begin{equation}
M_{Z}=2.09M_{\rm O}+1.06M_{\rm Fe}
\label{eq:met}
\end{equation}
according to the mixture of alpha (C, N, O, Ne, Mg, Si, S) and iron (Fe, Ni) group elements relevant for the sun \citep{Asplund2009}. The code accounts for metallicity dependent cooling by using tabulated cooling functions of \citet{sutherlanddopita93} for gas temperatures of $10^{4-8.5}$~K, and rates from \citet{rosenbregman95} for cooling down to lower temperatures. Heating from the ultraviolet background radiation is accounted for by using the model of \citet{haardtmadau96}, assuming a reionization redshift of $z=8.5$. Self-shielding is modelled following the model of \citet{AubertTeyssier2010}. Finally, we follow \citet{Agertz09b} and adopt an initial metallicity of $Z = 10^{-3} \Zsun$ in the high-resolution zoom-in region in order to account for enrichment from unresolved population III star formation \citep[e.g.][]{Wise2012}. 

\begin{figure*}
	\includegraphics[width=\linewidth]{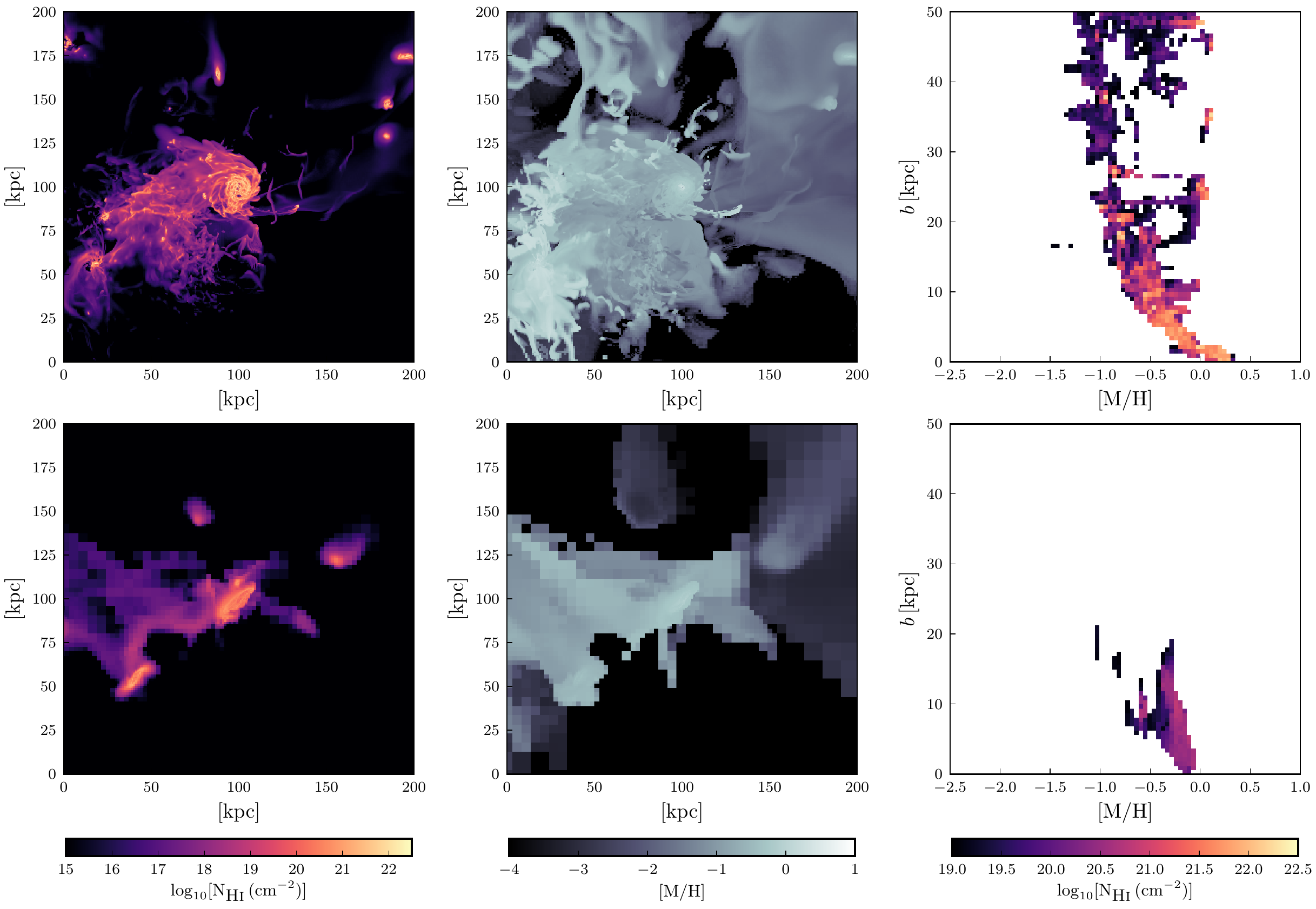}
    \caption{\ion{H}{i} column density projection (\textit{left}), neutral gas metallicity (\textit{center}), and projected separation as a function of metallicity for gas above the lower sub-DLA column density limit (\textit{right}). The top row features the simulated central galaxy at fiducial resolution (see Section \ref{sec:setup}), with the bottom row presenting a low resolution version, comparable to the typical  numerical resolution achieved in large volume cosmological simulations discussed in Section \ref{sec:fNH}. The enhanced resolution increases the damped \ion{H}{i} ($\log _{10}[\mathrm{N}_\textsc{Hi}~(\mathrm{cm}^{-2})] > 19$) covering fractions at large projected radii, and reveals a complex, turbulent disc-halo interface with small-scale \ion{H}{i}-structures.}
    \label{fig:exemplify}
\end{figure*}

\subsection{Model verification and analysis}
\label{sec:fNH}
By accounting for the above stellar feedback budget, \citet{AgertzKravtsov2015, agertzkravtsov2016} simulated the cosmological assembly of Milky Way-mass galaxies, demonstrating that it led to a realistic late-type galaxy matching observed properties such as disc size, the presence of a thin and thick stellar disc, stellar and gas surface density profiles, the Kennicutt-Schmidt relation, the stellar mass-gas metallicity relation (and its evolution), and a specific angular momentum typical of spiral galaxies of the Milky Way mass (stellar mass of $\approx 5\times 10^{10} \Msol$). 

The simulation considered in this work reaches a significantly higher mass and spatial resolution, and matches the above mentioned observational relations (to be presented in Agertz et al. in prep). Furthermore, we find that $\textsc{Hi}$ column density distributions, $f(N_\textsc{Hi})$, at all redshifts feature a turnover at column densities $N_{\textsc{Hi}}\sim 10^{21}~{\rm cm^{-2}}$, as observed both at high and low redshifts \citep[][]{Zwaan2005,Noterdaeme2009}, and in the ISM of individual local galaxies \citep[see analysis by][]{Erkal2012}. Our simulation does therefore not suffer from an excess of high column density $\textsc{Hi}$
($N_{\textsc{Hi}}\gg 10^{21}~{\rm cm^{-2}}$) in the inner few kpc of galaxies, as found by \citet[][]{Erkal2012} in their high resolution simulations with inefficient stellar feedback\footnote{Using simulations at lower numerical resolution, \citet[][]{Altay2011} found that the $\textsc{Hi}-{\rm H}_2$ transition was sufficient to explain the observed turnover, in contrast to \citet[][]{Erkal2012} \citep[see also][]{Altay2013} who argued that establishing a realistic star formation-feedback cycle is crucial for explaining the column density distribution of DLAs.}. As such, our zoom simulation is a relevant platform for studying the nature of DLA and sub-DLA LOS in and around massive disc galaxies. 

Using the dark matter halo finder HOP \citep[][]{HOP1998}, we identify all dark matter halos\footnote{Substructures are not robustly identified using HOP, but as we are mainly interested in \emph{central} galaxies, this is no issue.} within the simulated high resolution volume, in all simulation snapshots (see Section \ref{sec:setup}). For each halo, we generate two-dimensional maps of $N_\textsc{Hi}$ and metallicity ([M/H]). We obtain \textsc{Hi} from the neutral gas density in each cell by correcting for its molecular hydrogen content using the model by \citet[][]{Krumholz2008,Krumholz09}, as implemented in \cite{AgertzKravtsov2015}.

All maps cover an area of $200\times 200\kpc^2$, and are centered on the dark matter and stars using a shrinking sphere algorithm \citep[][]{Power2003}. The pixels have the same sizes as the finest cell sizes available in the simulation at the given snapshot, typically $\Delta x\sim 20\pc$. In each pixel, the quantities have been computed as 
\begin{equation}
N_\textsc{Hi}=\sum_i n_{\textsc{Hi},i}\Delta x_i
\end{equation}
and
\begin{equation}
{\rm [M/H]}= \frac{{\sum_i n_{\textsc{Hi},i}{\rm [M/H]}_i}}{\sum_i n_{\textsc{Hi},i}},
\end{equation}
where the sum runs over all cells along a given pixel LOS along the simulation volume (with a 200 kpc depth), $n_{\textsc{Hi},i}$ is the local cell density of neutral hydrogen, [M/H]$_i$ the local cell gas metallicity, and $\Delta x_i$ the size of a cell. For each snapshot we compute maps from the three Cartesian directions. In subsequent sections we will present results both for single snapshots and directions to illustrate our findings, as well as results averaged over all analysed directions and different redshift ranges. Finally, quoted impact parameters $b$ are defined as the projected distance from the center of the main galaxy to a point of interest. 

Numerical studies of DLAs have generally focused on large simulations volumes in order to capture a statistical sample of galaxies, at the sacrifice of numerical resolution (see Section \ref{sec:introduction}). Indeed, modern cosmological simulations of galaxy formation in domains with sizes of over tens of Mpc, aimed at reproducing the stellar luminosity function, e.g. {\small EAGLE} \citep[][]{Schaye2015} and Illustris \citep[][]{Vogelsberger2014}, have been restricted to spatial resolutions of $\sim 0.5-1\kpc$. In contrast,  we are here interested in a specific set of observations relating DLAs and sub-DLAs, at impact parameters up to $b\sim 50\kpc$, to confirmed host galaxies, here predominantly massive and star forming (see Section \ref{subsec:comparison}), making a comparison to high resolution zoom simulations more appropriate.

The importance of high numerical resolution is illustrated in Figure~\ref{fig:exemplify}. The top row shows $N_{\textsc{Hi}}$ and [M/H] maps, as well as $b$ vs.  [M/H], at $z=1$ for the most massive galaxy in our zoom region ($M_\star\sim 5\times 10^{10}\Msol$). The bottom row shows the same quantities, but for the simulation run at 32 times lower spatial resolution ($\Delta x\sim 640 \pc$), and dark matter particle masses of $1.4\times 10^6\Msol$, compatible with the aforementioned large volume simulations. While the low resolution simulation also leads to an extended disc, with a similar stellar mass at $z=0$ (made possible by the stellar feedback driven galactic outflows), the lack of detail in the circumgalactic medium (CGM) compared to the fiducial high resolution model is striking. The fiducial model resolves a greater number of low mass satellites, all found to contribute to DLA and sub-DLA detections (see also Section \ref{subsec:disc_vs_halo}). In addition, galaxy interactions, inflows and feedback driven outflows create complex gas structures with large covering fractions (see Sections \ref{subsec:disc_vs_halo} and  \ref{subsec:fcov_evolution}). A significant fraction of these gas structures show high levels of metal-enrichment for which feedback and satellite accretion is key; most prominently observed as potential damped \ion{H}{i} detections at $b>20\kpc$ in the $b-$[M/H] plane (Section \ref{subsec:comparison}). All of the above are crucial in order to interpret the observations presented in the next following analysis.

\section{Results}
\label{sec:results}

\subsection{Comparing simulations to analytical models}
\label{subsec:comparison}
\begin{figure*}
	\includegraphics[width=\linewidth]{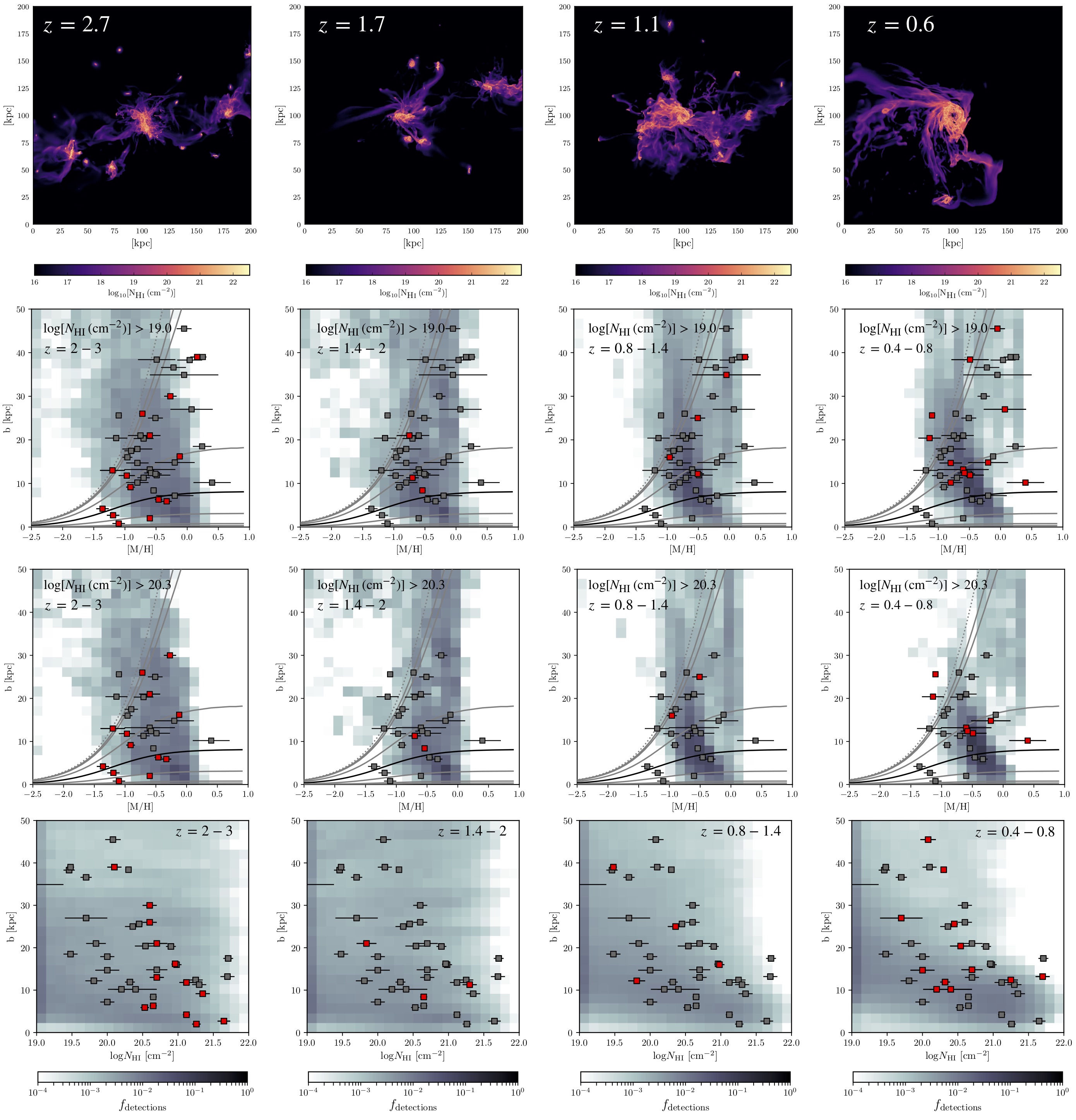}
\caption{\textit{First row:} \ion{H}{i} column-density projections of the simulated central galaxy. \textit{Second row:} Angle-averaged detection-fractions of sight-lines above the sub-DLA column-density limit ($\log _{10}[\mathrm{N}_\textsc{Hi}~(\mathrm{cm}^{-2})] = 19$) within an observationally motivated beam of 50 kpc radius (in projected separation). \textit{Third row:} Same a second row, cut at the DLA column-density limit ($\log _{10}[\mathrm{N}_\textsc{Hi}~(\mathrm{cm}^{-2})] = 20.3$). \textit{Fourth row:} Angle-averaged detection-fractions in the space spanned by projected distance and \ion{H}{i} column-density. Image columns depict the redshift-evolution in the relations, moving from high (\textit{left column}) to progressively lower redshifts. The MC19 compilation of observations has been overplotted as individual data-points. The data has been filtered to the column-density threshold of the row, and is colour-coded to depict the data in the redshift-range of the image column in red. In row two and three, we have overplotted the mean (black) and $1\sigma$-, $2\sigma$- and $3\sigma$-contours (grey) lines of the analytical F08+K17 model.}
    \label{fig:model_comparison}
\end{figure*}

By combining scaling relations, the simple analytical F08+K17 model was constructed, which successfully describes DLAs as resulting from random LOS through inclined \ion{H}{i} discs of the same population of galaxies traced by the Lyman-break technique (LBGs)  \citep{Fynbo2008, Krogager2017}. This model combines a $z\sim3$ UV luminosity function \citep{Reddy2008}; a $z\sim 3$ metallicity-luminosity relation \citep{Pettini2001}; a Holmberg relation to describe the size of the \ion{H}{i}-disc as a function of the galaxy luminosity \citep[following][]{Wolfe1986}; and a prescription for the metallicity gradient. In light of its success, with our simulations we now ask if the assumed scaling relations are reproduced in the simulations and if so at which redshifts.

To answer this, we use HOP to identify the most massive dark matter halo, and follow its evolution in redshift from $z = 3$ to $z = 0.4$. This halo grows in stellar mass from \mbox{$9.6 \leq \log _{10}[\mathrm{M}_\star~(\mathrm{M}_\odot)] \leq 10.8$}, which matches the LBG population stellar mass distribution \mbox{$\log _{10} [\mathrm{M}_\star~(\mathrm{M}_\odot)] = 9.87 \pm 0.53$} \citep{Reddy2009} from whose luminosity function the F08+K17 model was constructed. We compute $b~(\mathrm{kpc})$, $[\mathrm{M/H}]$, and $\log _{10} [N_\textsc{Hi}~(\mathrm{cm^{-2})}]$ at each redshift and along each side of the simulation-box using the method described in Section \ref{sec:fNH}. Binning in redshift intervals, we calculate the angle-averaged detection-fractions ($f_\mathrm{detections}$) at each locus. The results are presented in Fig~\ref{fig:model_comparison} and consists of probability-functions in the $b~(\mathrm{kpc})-[\mathrm{M/H}]$ -and the $b~(\mathrm{kpc}) - \log _{10} [N_\textsc{Hi}~(\mathrm{cm^{-2})}]$ parameter space, with $f_\mathrm{detections}$ normalized for each panel. We show these results together with predictions from the F08+K17 model (with the mean relation in black, and $1\sigma$-; $2\sigma$- and $3\sigma$ intrinsic dispersions (grey) contours), and the MC19 compilation for a sequence of redshifts (columns). Row one shows a simulation snapshot centered on the galaxy; rows two- and three show the space spanned by $b~(\mathrm{kpc})-[\mathrm{M/H}]$ with a cut on the column-density at the sub-DLA and at the DLA lower limits, respectively; and row four spans the $b~(\mathrm{kpc}) - \log _{10} [N_\textsc{Hi}~(\mathrm{cm^{-2})}]$ parameter space. 

Figure \ref{fig:model_comparison} reveals significant detection probabilities of LOS above the DLA column density threshold extending to impact parameters of $50~\mathrm{kpc}$ from the host center, across all redshifts. However, the distribution clearly changes with redshift. Qualitatively, this can be understood as follows; due to the cosmic density evolution with redshift, \mbox{$\rho (z) \propto (1+z)^3$}, at higher redshifts the higher density environments gives rise to more interactions and more tidal and/or turbulent disturbances. At $z>2$, this inhibits gas from settling into coherent and long-lived structures such as extended \ion{H}{i} discs. We therefore see $f_\mathrm{detections}$ across a broad range of grid-points in the $b~(\mathrm{kpc})-[\mathrm{M/H}]$ -and the $b~(\mathrm{kpc}) - \log _{10} [N_\textsc{Hi}~(\mathrm{cm^{-2})}]$ parameter spaces alike. This is particularly clear in the highest redshift panel ($z=2-3$, Fig~\ref{fig:model_comparison} \textit{bottom left}), for which our simulations generate detections across the whole parameter-space. 

Towards progressively lower redshift, merger-rates and densities drop, and the \ion{H}{i} gas is allowed to settle. This includes the formation of an \ion{H}{i} disc which encodes information on a metallicity-gradient, most prominently observed as the diagonal stripe of high detection-probability (Fig. \ref{fig:model_comparison} \textit{third row, rightmost panel}, but see also Section \ref{subsec:disc_vs_halo} for a direct observation of the disc in this parameter-space). In the same row, we also find a region of diffuse (lower detection probability per unit grid-point) $f_\mathrm{detections}$ but extending to large impact parameters. This is indicative of a significant contribution from the circum-galactic medium and the extended halo to the DLA cross-section (see Sections \ref{subsec:disc_vs_halo} and \ref{subsec:b_evolution}).

\begin{table}
 \begin{center}
 \caption{Simulation-based detection fractions of DLA-LOS enclosed within the F08+K17 model $1\sigma$-, $2\sigma$-, and $3\sigma$-confidence regions.}
 \label{tab:compare1}
 \begin{tabular}{cccc}
  \hline
  Redshift range & $1\sigma$ & $2\sigma$ & $3\sigma$ \\
   				 & $(0.68)$  & $(0.95)$  & $(0.99)$ \\
  \hline
  $3.0-2.0$ & $0.41$ & $0.74$ & $0.77$ \\
  $2.0-1.4$ & $0.41$ & $0.81$ & $0.85$ \\
  $1.4-0.8$ & $0.58$ & $0.89$ & $0.91$ \\
  $0.8-0.4$ & $0.78$ & $0.97$ & $0.98$ \\
  \hline
 \end{tabular}
 \end{center}
\end{table}

In Table~\ref{tab:compare1} we quantify the agreement between the F08+K17 model and our simulation by calculating the summed detection-fraction enclosed within the $1\sigma$- $2\sigma$- and $3\sigma$ limits. The DLA LOS detection fractions in simulations converge to model predictions towards progressively lower redshifts, reaching a good agreement at $z\sim 0.8$.

Relaxing the threshold on the column-density to include detection-fractions from sub-DLA LOS in $b~(\mathrm{kpc})-[\mathrm{M/H}]$ (Figure \ref{fig:model_comparison}, \textit{second row}), we find that $f_\mathrm{detections}$ becomes less concentrated and covers a larger region in the parameter-space at all redshifts. Phenomenologically, this suggests that sub-DLA LOS fill the void between islands of DLA-$f_\mathrm{detections}$, and could be related to a density gradient such that sub-DLAs on average trace lower-density gas. We pursue this by separating the galaxy into a halo and a disc (Section \ref{subsec:disc_vs_halo}) and quantify statistical differences between the average DLA and sub-DLA LOS in Sections \ref{subsec:b_evolution} and \ref{subsec:fcov_evolution}. 

In Table \ref{tab:compare2} we demonstrate how the inclusion of sub-DLA column densities affects the match between simulation $f_\mathrm{detections}$ and the analytical disc model predictions. We find that such an inclusion results in a worse match at all redshifts. This most probably reflects the underlying Holmberg relation in the construction of the model, which works well down to the lower limit of DLA column densities. Indeed, this statistical discrepancy also matches the observational results of \cite{Rhodin2018}, who found an excess of sub-DLAs beyond the predicted $1\sigma$ model region. We therefore conclude that in order to match simulations and observational data below the formal DLA limit, the analytical disc-model must be amended or expanded.

\begin{table}
 \begin{center}
 \caption{Same as Table \ref{tab:compare1}, but including sub-DLA LOS.}
 \label{tab:compare2}
 \begin{tabular}{cccc}
  \hline
  Redshift range & $1\sigma$ & $2\sigma$ & $3\sigma$ \\
   				 & $(0.68)$  & $(0.95)$  & $(0.99)$ \\
  \hline
  $3.0-2.0$ & $0.41$ & $0.70$ & $0.74$ \\
  $2.0-1.4$ & $0.39$ & $0.76$ & $0.81$ \\
  $1.4-0.8$ & $0.42$ & $0.77$ & $0.81$ \\
  $0.8-0.4$ & $0.54$ & $0.84$ & $0.87$ \\
  \hline
 \end{tabular}
 \end{center}
\end{table}

\subsection{The fraction of DLA/sub-DLA systems in discs and halos}
\label{subsec:disc_vs_halo}

\begin{figure*}
	\includegraphics[width=\linewidth]{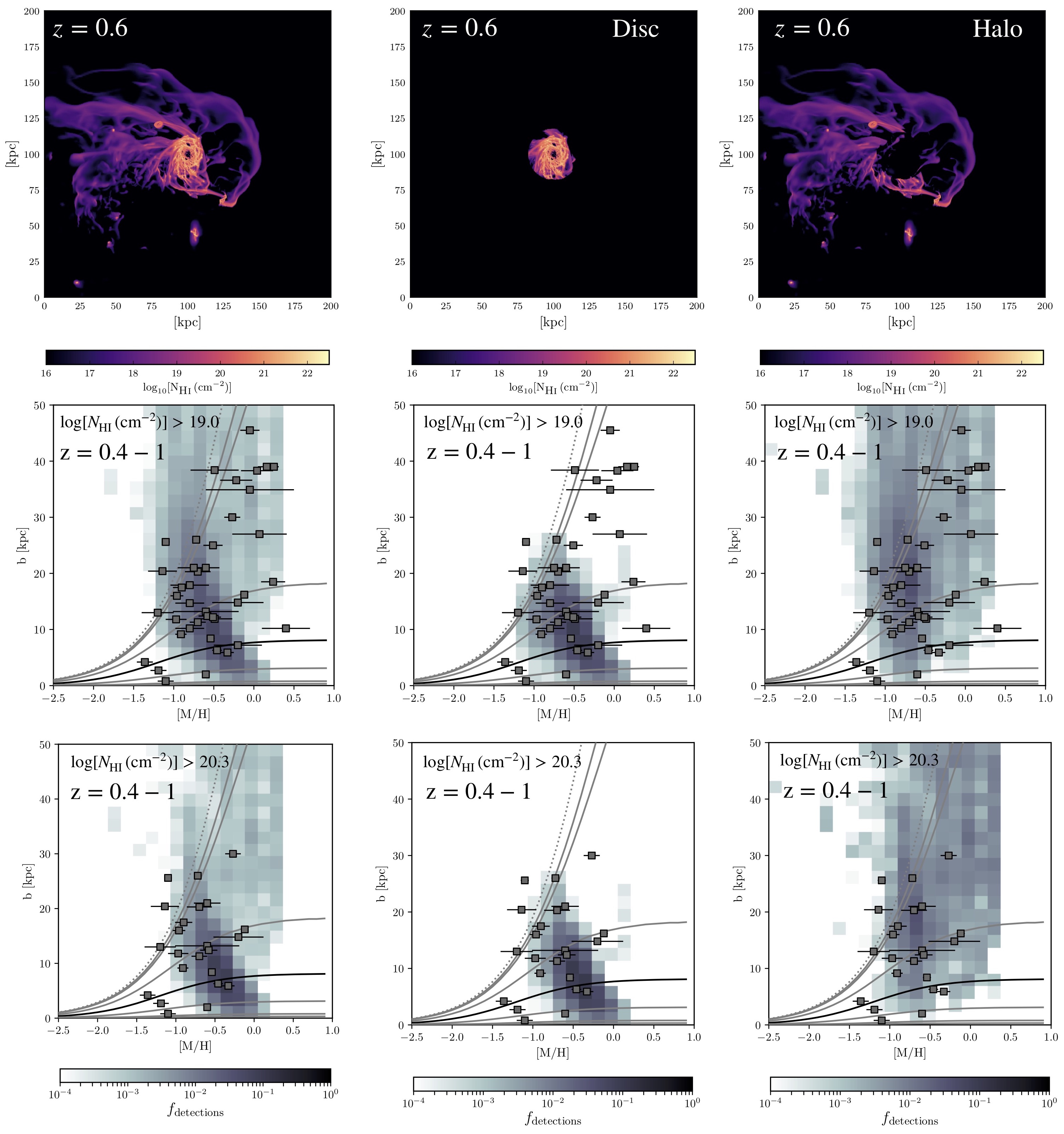}
    \caption{\ion{H}{i} column density projection of the central halo (\textit{top}) split into three panels depicting the total \ion{H}{i} gas projection (\textit{left}), the disc component (\textit{center}) and halo (\textit{right}). The bottom panels display the associated distributions of detection-fractions in the $b-[\mathrm{M/H}]$ plane, with cuts at the sub-DLA and DLA threshold, respectively. The integral of detection-fractions in a panel measures the contribution of that structure to the total number of LOS producing damped \ion{H}{i} absorption above the associated column-density threshold.}
    \label{fig:dischalosep}
\end{figure*}

The relative contributions of interstellar  (``disc'')  and circum-galactic  (``halo'') gas to DLA and sub-DLA detections are not yet known. Despite employing an infinitely thin circular disc to define a geometry on which to distribute the DLA cross-section, the F08+K17 model does not claim to describe galactic \ion{H}{i} discs. In simulations, such a separation is complicated as these gaseous components are ill-defined, especially for the earliest stages of disc formation. This is especially true for gas in the interface between the disc and the halo, related to accretion and outflows. The results below should hence be treated as indicative.

We adopt the following simple approach for separating neutral disc gas from halo gas, and apply it to our most massive galaxy in the zoom-in region. For each simulation snapshot, the angular momentum vector is computed for all baryons residing within $3 r_{1/2}$ of the central galaxy, where $r_{1/2}$ is the (baryonic) half mass radius. This defines a disc plane, for which we compute the density profile $n(r)$ of \emph{neutral gas}, where $r$ is the spherical radius. Such profiles are nearly exponential, with a fall-off below $n(r)\sim 10^{-2}\,{\rm cm}^{-3}$ due to the inability of gas to self-shield against the background UV field \citep[also found in the radiative transfer calculations by][]{AubertTeyssier2010}. We define the size of the disc to be the radius $r_{\rm d}$ where $n(r)$ falls below $n=10^{-3}\,{\rm cm}^{-3}$, but note that our conclusions are not very sensitive to cut-off densities in the range to $n=10^{-4}-10^{-2}\,{\rm cm}^{-3}$. At $z<1$, $r_{\rm d}$ fall in the range $\sim 15-30$ kpc. For each simulation snapshot, all neutral gas inside of the cylindrical radius $r_{\rm d}$ and below an altitude above the disc plane of $h=8$ kpc is defined as 'disc', and the gas outside of this domain is defined as 'halo'. 

At high redshifts ($z\gtrsim 1-2$), mergers and and galaxy interactions yield mostly transients discs - if at all. To allow for a cleaner disc-halo separation, we restrict our analysis to $z<1$, and in order to convey that a robust (as opposed to a transient) disc is identified, we stack the results between $z=0.4-1$. In this way, we ensure that our methodology is not biased to the conditions of a single snapshot. The results of the component separation are shown in in Figure \ref{fig:dischalosep}, for which the top row displays an \ion{H}{i} column density map of the galaxy (\textit{left}), the \ion{H}{i} disc (\textit{center}) and halo (\textit{right}) at $z=0.6$. Row two and three display the associated detection fractions in the $b-$[M/H] plane, with lower column-density thresholds set to sub-DLA and DLA limits, respectively. Our simple approach allows a clear disc component to be extracted, with the residual halo gas containing satellites, stripped filamentary gas, and gas at smaller impact parameters residing just outside, but soon to be accreted onto, the disc.

The disc component is visually distinct in the $b-$[M/H] relation, both for DLA LOS alone, and when sub-DLA LOS are included, and extends all the way to the $3\sigma$ confidence relation of the analytical F08+K17 model. In fact, $99.6\%$($88\%$) of all simulated DLA sightlines lie within the $2$(1)$\sigma$ relation, indicating that the F08+K17 model does not necessarily represent the physical nature of turbulent discs, but illustrates why this model has been highly successful in capturing observational properties of DLAs \citep[][]{Krogager2017}. For the above disc definition, $\sim 40\%$ ($60\%$) of LOS with $\log _{10}[\mathrm{N}_{\textsc{Hi}}~(\mathrm{cm}^2)]>19$ intersect 'halo' ('disc') gas, and  $\sim 20\%$ ($80\%$) for $\log _{10}[\mathrm{N}_{\textsc{Hi}}~(\mathrm{cm}^2)]>20.3$. By allowing for a disc definition with $h=4$ kpc, rather than 8 kpc, we find identical results, indicating that our results are robust to reasonable parameter changes. In summary, our experiments illustrate that for the extended disc galaxy of the type analysed here, compatible with $L_\star$ galaxies ($M_\star\sim {\rm few}\times 10^{10}\Msol$), DLAs predominantly originate in extended \ion{H}{i} discs, but with halo gas playing a role at all impact parameters - especially at high impact parameters and in the sub-DLA regime.

Both simulations and observations indicate a significant amount of metal rich DLA/sub-DLA LOS at high impact parameters ($b\gtrsim 25$ kpc). From the simulated component separation, it is clear that most of the gas contributing to these LOS is of halo origin. This can result from a high metal column, or a low \ion{H}{i} column. We interpret this as a clear sign of metal rich gas stripped from dwarf satellites, as the shallow potential wells in these galaxies allow for efficient unbinding of enriched gas via outflows, ram pressure and tidal stripping. This scenario is supported by Figure \ref{fig:exemplify} which shows that the metallicity map does not necessarily follow the \ion{H}{i} map but instead shows higher metal content extending from the satellites; and visually by the presence of satellites and stripped tidal tails in the halo component of Figure \ref{fig:dischalosep}.

\subsection{A redshift evolution in the mean impact parameters}
\label{subsec:b_evolution}

\begin{figure}
	\includegraphics[width=\columnwidth]{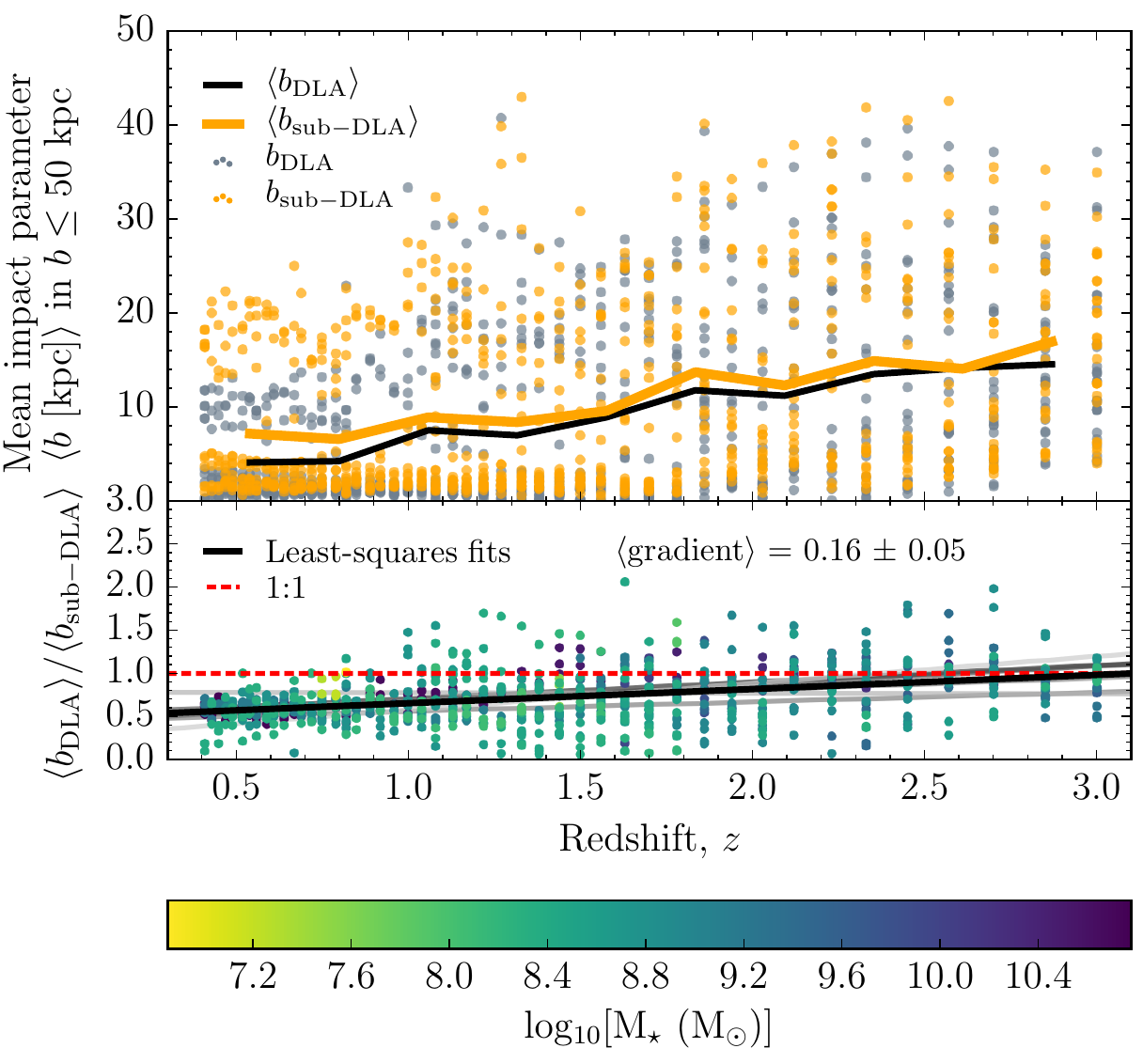}
    \caption{Angle-averaged mean impact parameters ($b~\mathrm{[kpc]}$) of sub-DLA and DLA sight-lines in seven halos followed in our simulations. \textit{Top panel:} Mean impact parameter of DLA (grey) and sub-DLA (orange) LOS of individual snapshots. Binned mean trends for DLA and sub-DLAs impact parameters as a function of redshift are depicted as solid lines following the same colour-coding. \textit{Bottom panel:} Data represents the mean DLA to sub-DLA impact parameter ratio for each snapshot and from each side of the simulation box, colour-coded to stellar mass. The dashed red line marks a one to one ratio. Solid grey lines show fits to separate halos. The solid black line depicts the mean relation.}
    \label{fig:b_z}
\end{figure}

The results of Section \ref{subsec:comparison} suggest that there is a redshift evolution in the internal distribution of sub-DLA and DLA LOS through a galaxy, and that this evolution may be different for sub-DLAs and DLAs. We know from observations that damped \ion{H}{i} absorbers follow a column-density dependent distribution of impact parameters \citep[][]{Christensen2007, Monier2009, Fynbo2010, Fynbo2011, Meiring2011, Rao2011, Rahmani2016}. Marginalizing over redshifts, these observations state that DLA LOS are found systematically closer to their host galaxy than sub-DLA LOS by a factor of $\sim 2$ \citep{Rao2011,Rhodin2018}. Likewise, investigations of the $\Delta V_{90}-[\mathrm{M/H}]_\mathrm{abs}$ relation \citep[e.g.][]{Ledoux2006,Moller2013,Neeleman2013} have shown that DLAs have smaller mean velocity widths and lower metallicities than sub-DLAs \citep[e.g.][]{Som2015}. Taken together, these observations suggest that the two classes of absorbers on average trace different parts of their hosts.

We now explore whether we can quantify such average differences. The simplest non-parametric statistic to quantify the difference is the mean impact parameters of sub-DLA and DLA LOS. From here onwards, these will be referred to as $\langle b_\mathrm{sub-DLA} \rangle$ and $\langle b_\mathrm{DLA} \rangle$ respectively. These quantities are particularly interesting from an observational point of view, as impact parameters are amongst the simplest observables and readily available for 43 absorber-galaxy pairs in the MC19 compilation. In addition, the analysis presented in Section \ref{subsec:comparison} suggest that DLA LOS can be found at large impact parameters with significant probabilities.

We first calculate $\langle b_\mathrm{sub-DLA} \rangle$ and $\langle b_\mathrm{DLA} \rangle$. To ensure that the galaxy analysed in previous sections is not an outlier, we allow HOP (see Section \ref{sec:fNH}) to identify the seven most massive halos. This allows us to cover diverse galaxy environments, a stellar mass range of $6.7 < \log _{10} [\mathrm{M}_\star (\mathrm{M}_\odot)] < 10.8$ which covers most observations, and to span a redshift range $0.4<z<3.0$. Using the information on $\langle b_\mathrm{sub-DLA} \rangle$ and $\langle b_\mathrm{DLA} \rangle$ together with the associated redshift and stellar mass from the three Cartesian directions and for each of the halos, we present the results in Figure \ref{fig:b_z}. The top panel displays the individual data of $\langle b_\mathrm{sub-DLA} \rangle$ (orange) and $\langle b_\mathrm{DLA} \rangle$ (grey) together with their binned-mean relations in the respective colours, as a function of redshift. The results indicate a co-evolution in mean DLA -and sub-DLA impact parameters towards progressively smaller values with time. However, perhaps the most striking feature is the large scatter in the mean values at any redshift. The scatter is set from the range of galaxy-sizes spanned by the seven halos, and the effect of a fixed $50~\kpc$ beam. To cancel the size-dependence and enhance any differential change in the mean impact parameters with redshift, in the bottom panel, we therefore proceed to plot the ratio of the means; $\langle b_\mathrm{DLA} \rangle / \langle b_\mathrm{sub-DLA} \rangle$. A horizontal line implies a scale-free fraction, which at unity translates to identical mean impact parameter sub-DLA and DLA LOS. For reference, we display such a $1:1$ relation as a red dashed line.

To ensure that any correlation in $\langle b_\mathrm{DLA} \rangle / \langle b_\mathrm{sub-DLA} \rangle$ with redshift is indeed real, and not a consequence of an underlying correlation with stellar mass, we calculate the Pearson (PC) coefficient, the Spearman rank-order correlation (SC) coefficient, and their associated null-hypothesis's p-values. Expressed \mbox{$\mathrm{Test}~(\mathrm{parameter1}, ~\mathrm{parameter2}) = [\mathrm{coefficient},~ \mathrm{p-value}]$} we find \mbox{$\mathrm{PC}(\langle b_\mathrm{DLA} \rangle / \langle b_\mathrm{sub-DLA} \rangle, ~\log _{10} [\mathrm{M}_\star (\mathrm{M}_\odot)]) = [0.073, \sim 10^{-2}]$} and \mbox{$\mathrm{SC}(\langle b_\mathrm{DLA} \rangle / \langle b_\mathrm{sub-DLA} \rangle, ~\log _{10} [\mathrm{M}_\star (\mathrm{M}_\odot)]) = [0.144, \sim 10^{-5}]$}, whereas significantly higher correlation coefficients are identified for \mbox{$\mathrm{PC}(\langle b_\mathrm{DLA} \rangle / \langle b_\mathrm{sub-DLA} \rangle, ~z) = [0.396, \sim 10^{-31}]$}, \mbox{$\mathrm{SC}(\langle b_\mathrm{DLA} \rangle / \langle b_\mathrm{sub-DLA} \rangle, ~z) = [0.399, \sim 10^{-32}]$}. With consistent results from a Pearson and a Spearman test, both strongly suggesting that redshift is indeed the primary driver in the impact parameter evolution,  we proceed to ask whether it is driven by a particular halo.

We find consistent correlation coefficients of the same magnitude using both the Pearson and the Spearman test (six positive and one negative correlation) for the individual halos. The magnitude of these coefficients for individual halos are similar to that of the population as a whole, but at elevated p-values in a broad range $\sim [10^{-33} : 10^{-1}]$. We therefore proceed to explicitly calculate linear fits to individual halos (Figure \ref{fig:b_z}, grey lines). These fits yield similar gradients that systematically favor an evolution in the mean impact parameter ratio with time. With a fit to all data, we find a mean gradient of $0.16 \pm 0.05$ per unit redshift, i.e. at a formal $3.2\sigma$ significance (Figure \ref{fig:b_z}, black solid line).

This mean evolution in $\langle b_\mathrm{DLA} \rangle / \langle b_\mathrm{sub-DLA} \rangle$-ratio with redshift is remarkably consistent with the analysis presented in Section \ref{subsec:comparison}. At $z\sim 3$ the ratio intercepts the 1:1 correlation. This is in part caused by the \textit{presence} of multiple galaxies within the $50~\mathrm{kpc}$ projected radius beam - each one generating its own internal DLA and sub-DLA LOSs - which on average will tend to average out any difference between the mean impact parameters discovered in the beam. In part, it is caused by the \textit{interaction} between galaxies in the beam, which causes irregular gas flows that inhibit the formation of coherent structures, rendering a stochastic distribution. Moving down in redshift, the distribution of sub-DLA and DLA LOS diverge such that, on average, they are separated by a factor of two below redshifts $z\sim 0.5$.

It is striking that such a factor of $\sim 2$ difference in mean projected separation between sub-DLA and DLA LOS is retrieved in observations at redshifts $z < 1.0$ \citep{Rao2011, Rhodin2018}. Indeed, even the absolute mean impact parameters derived in these studies ($\langle b\rangle _\mathrm{DLA}^\mathrm{empirical} \sim 15~\kpc$, $\langle b \rangle _\mathrm{sub-DLA}^\mathrm{empirical} \sim 30~\kpc$) are marginally consistent with some of the high means retrieved for individual galaxies in our simulations at comparable redshifts (see Figure \ref{fig:b_z}, top panel). As per Section \ref{subsec:comparison} and Figure \ref{fig:model_comparison} bottom row, however, we find significant detection fractions of both sub-DLA and DLA LOS to projected separations $\sim 50~\kpc$ at all redshifts, and marginal inconsistencies in the absolute mean values between simulations and observations may reflect (observational) low number statistics and that cosmological zoom simulations have not converged on the spatial scales required to resolve cold gas in the CGM \citep{McCourt2018, Hummels2018}. We also note that \citet[R14]{Rahmati2014} identify an anti-correlation between median impact parameter and $N_\textsc{Hi}$, for which they report $\langle b \rangle _\mathrm{sub-DLA}^\mathrm{R14} \sim 7 ~\mathrm{pkpc}$ and $\langle b \rangle _\mathrm{DLA}^\mathrm{R14} \sim 2~\mathrm{pkpc}$. Albeit a large intrinsic scatter in the data, with 15-85 per cent percentiles which we estimate to extend from $2-29~\mathrm{kpc}$ and $0.5 - 4~\mathrm{kpc}$ for sub-DLAs and DLAs, respectively \cite[][private communication]{Rahmati2014}, their median values are significantly lower than those identified in observations and in our simulations. Additionally, despite a consistent factor of two difference in the means, their analysis is based on redshift $z=3$, whereas our analysis of the redshift evolution suggests that such a clear separation is only in place at $z \lesssim 0.7$.

\subsection{Do sub-DLA and DLA sight-lines probe galaxies of different mass?}
\label{subsec:fcov_evolution}

An alternative explanation to the idea that sub-DLA and DLA LOS trace different relations \textit{within} a galaxy, is that they on average probe intrinsically \textit{different} galaxy populations; sub-DLA LOS on average probing more massive galaxies than do DLA LOS \citep[e.g.][]{Khare2007,Kulkarni2010}. Similar conclusions were also reached with deep ground-based imaging of sub-DLA hosts, which suggest that these systems preferentially sample more luminous - and therefore more massive galaxies than DLAs \citep{Meiring2011}. By measuring the stellar mass of the absorbing galaxies from matching spectral energy distribution galaxy templates to photometry, \cite{Augustin2018} and \cite{Rhodin2018} directly confirmed the apparent anti-correlation between the host's stellar mass and the \ion{H}{i} column-density of the absorber.

If real, this correlation is remarkable as we expect both high- and low column-density LOS to pass through a given galaxy. In addition, the shape of the column-density frequency distribution function ($f(N_\textsc{Hi}, X)$) forces the frequency of sub-DLAs to outnumber that of DLAs. From \citet[][P14]{Prochaska2014} and with integration limits $[19.00, 20.30]_\mathrm{sub-DLA}$ and $[20.30, 25.00]_\mathrm{DLA}$, one finds $N_\mathrm{DLA}/N_\mathrm{sub-DLA} \sim 0.38$, which together with the apparent correlation above implies an excess of massive gas-rich galaxies relative to low-mass galaxies, at odds with naive expectations from galaxy luminosity functions. A viable alternative is that the correlation is driven by sample selection effects. 

\begin{figure}
	\includegraphics[width=\columnwidth]{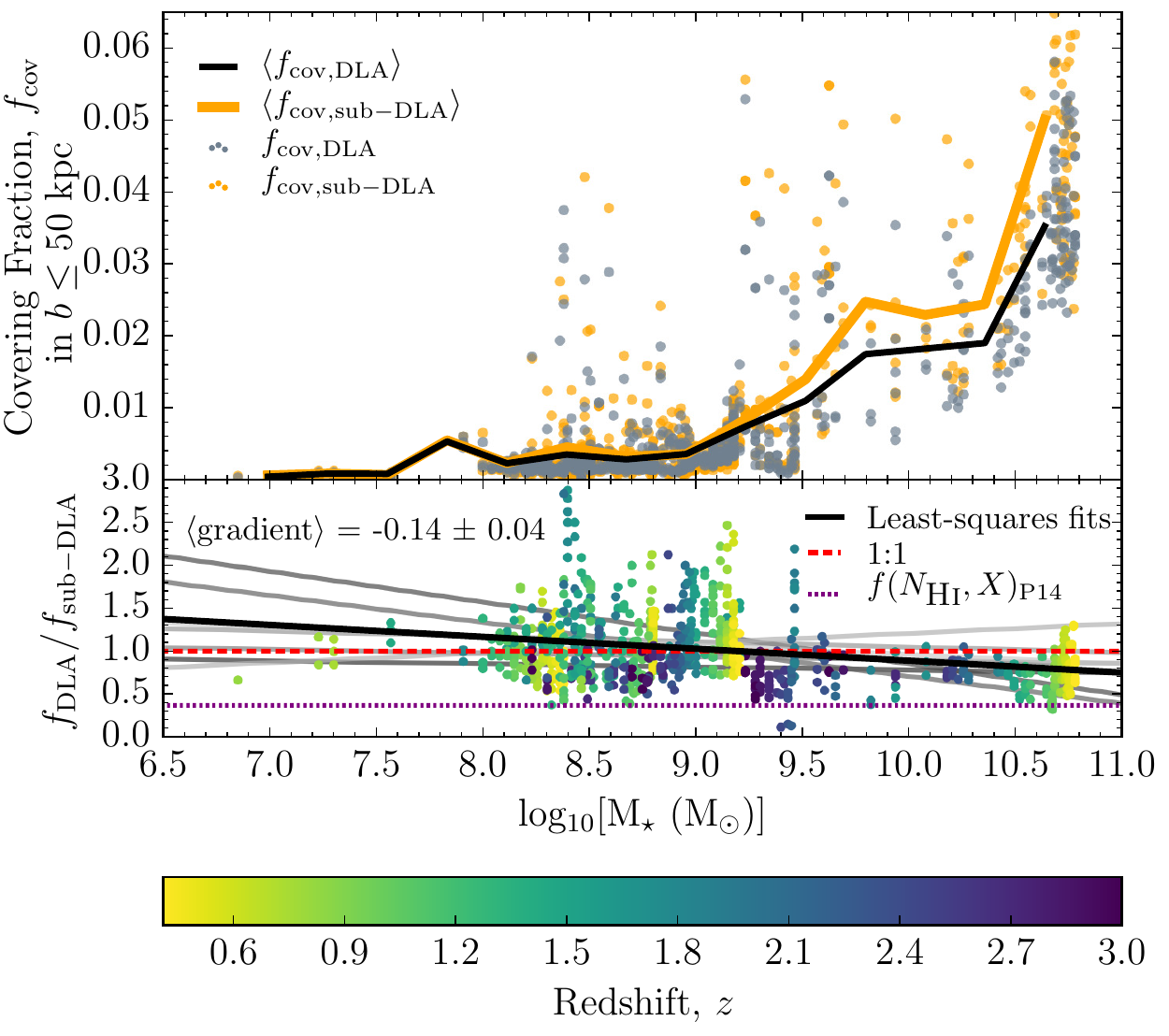}
    \caption{Same as Figure \ref{fig:b_z}, but for covering-fractions ($f_\mathrm{cov}$) of sub-DLA and DLA LOS as a function of stellar mass. The \textit{bottom panel} additionally includes the expected cosmological ratio of the frequency in DLA to sub-DLA LOS based on the P14 column-density distribution function (dotted purple line, see text for more details).}
    \label{fig:fcov_Mstar}
\end{figure}

Indeed \citet{Dessauges-Zavadsky2009} argue for a metallicity bias at low-redshifts, while at $z>1.7$ DLA and sub-DLA selections show statistically consistent metallicity distributions. Recent work on dust attenuation and quasar surveys confirms that we expect a dust-bias to preferentially act on high $\mathrm{N}_\textsc{Hi}$, massive, metal-rich, dusty galaxies selected against quasar LOS \citep{Vladilo2005,Zafar2013b,Fynbo2013b,Noterdaeme2015,Krogager2015,Zafar2015,Krogager2016,Fynbo2017}. In \citet{Rhodin2018} we argue that the detection void in the low mass, low $\ion{H}{i}$ column-density parameter space could be explained by a functional dependence of the ratio in covering fraction of sub-DLA to DLA LOS with stellar mass, such that sub-DLAs are preferentially detected in more massive galaxies.

To test this prediction, we calculate the covering fraction ($f_\mathrm{cov}$) of sub-DLA and DLA LOSs in each halo, and repeat the statistical analysis described in Section \ref{subsec:b_evolution}. We present the angle-averaged results in Figure \ref{fig:fcov_Mstar}. In the top panel, we plot the data for individual galaxies $f_\mathrm{cov}$ in DLA (grey) and sub-DLA (orange) LOS, together with binned-mean relations in the respective colours, as a function of stellar mass. $f_\mathrm{cov}$ for sub-DLA and DLA LOS correlate with stellar mass such that larger stellar masses, on average, are associated with larger $f_\mathrm{cov}$. At stellar mass $\log _{10}(\mathrm{M}_\star~[\mathrm{M}_\odot]) \sim 10$ the entire central region sees an increase in neutral gas, which leads to a rapid increase in the final covering fractions. For the most massive disc galaxy in our sample, this increase correlates with rapid disc growth at $z\sim 1.5$. This increase in gas disc sizes of galaxies forming in Miky Way-mass halos, around this redshift and stellar mass, is found in larger samples of zoom simulations (Kretschmer et al. in prep), as well as observationally in terms of the stellar size-mass relation \citep[see compilation by][]{AgertzKravtsov2015}.

Calculating the Pearson correlation coefficient; the Spearman rank-order correlation coefficient, and their associated null-hypothesis's p-value, we find \mbox{$\mathrm{PC}(z, f_\mathrm{DLA}/f_\mathrm{sub-DLA}) = [-0.162, \sim 10^{-6}]$} and \mbox{$\mathrm{SC}(z, f_\mathrm{DLA}/f_\mathrm{sub-DLA}) = [-0.164, \sim 10^{-6}]$}, whereas \mbox{$\mathrm{PC}(\log _{10}(\mathrm{M}_\star~[\mathrm{M}_\odot]), f_\mathrm{DLA}/f_\mathrm{sub-DLA}) = [-0.273, \sim 10^{-15}]$} and \mbox{$\mathrm{SC}(\log _{10}(\mathrm{M}_\star~[\mathrm{M}_\odot]), f_\mathrm{DLA}/f_\mathrm{sub-DLA}) = [-0.244, \sim 10^{-12}]$}. With consistent results, we conclude that a correlation in differential covering fraction with stellar mass indeed dominates over a correlation with redshift by a factor of $\sim 2$. In Figure \ref{fig:fcov_Mstar} lower panel, we show the ratio of DLA to sub-DLA covering fractions to envisage the differential scaling with stellar mass, and colour-code the data by redshift.

Having established the subordinate role of redshift to stellar mass in the covering fraction ratio, we proceed to ask whether the correlation is driven by a particular halo. Individual halos display similar correlation coefficients to the that of the population as a whole, but at elevated p-values in the range $\sim [10^{-3}:10^{-1}]$, suggesting that each halo contributes to the final p-value and correlation-strength of the population. We therefore proceed to calculate linear fits to the data of individual halos (overplotted as grey lines). With consistent trends, we find a mean differential covering fraction ratio per unit logarithmic stellar mass of $-0.14 \pm 0.04$. This corresponds to a weak functional dependence at a $3.5 \sigma$ significance which we over-plot as a black solid line. Qualitatively, this correlation with stellar mass is consistent with the results of \cite{FaucherGiguere2015} who report that the covering fraction of gas at different column-density thresholds is sensitive to individual dark matter halo masses for LBGs at $z\sim 2$.

Despite weak, if real and acting across the three decades in stellar mass $\sim [8.0 < \mathrm{M}_\star ~[\mathrm{M}_\odot] < 11.]$ identified in observations, we expect this gradient to account for $3 \times 0.14 \sim 0.4$, i.e. a $40 \%$ measured effect. We consider a more realistic value to be that produced within the $1\sigma$ dispersion in the stellar mass distribution of confirmed hosts. For the sample used in \citet{Rhodin2018}, with mean and dispersion in stellar mass of $9.87\pm 0.66$, we expect a measured effect of $\sim 20 \%$. Whereas we expect this effect to contribute to the observed anti-correlation, a direct comparison to the current set of data is beyond the scope of this work, as such a comparison has to account for sample selection functions and observational strategy bias.

As a reference-point, we instead calculate the predicted ratio of the number of DLAs to sub-DLAs, based on $f(N_\textsc{Hi}, X)$. With the P14 parametrization, $f(N_\textsc{Hi}, X)_\mathrm{P14}$, we expect a ratio $N_\mathrm{DLA}/N_\mathrm{sub-DLA} = 0.38$. We over-plot this estimator as a dotted purple line. Using the $f(N)$ statistic as an unbiased proxy for the cosmic average covering fraction ratio in DLAs to sub-DLAs, we find an over-density of DLA (or equivalently, an under-density of sub-DLA) LOS within a 50 kpc projected radius beam from the mass center of the hosting halo. From this, we conclude that the majority of sub-DLA are located at separations $b>50~\mathrm{kpc}$.

\section{Conclusions}
\label{sec:conclusion}

In this work, we have used a cosmological zoom-in hydrodynamics simulation of the assembly of a Milky Way mass galaxy to study beams of $50~\mathrm{kpc}$ radius centered on galaxies to study the environments capable of producing sub-DLA and DLA LOS. Building probability functions of observables and logging their redshift evolution, we have matched these to the most up-to-date compilation of spectroscopically confirmed absorbing galaxies, and to analytic models. The main results can be summarised as follows:

\begin{itemize}

    \item In agreement with recent results published by other groups, we find that increased numerical resolution leads to an excess of \ion{H}{i} coverage in galaxy halos. This increases the mean impact parameters of damped \ion{H}{i} absorbers and produces detection probabilities at all observed values.
    
    \item High numerical resolution and an effective feedback prescription allow us to reproduce the column density distribution function of blind quasar surveys and for individual galaxies. This validates the application of these simulations to study sub-DLA and DLAs in a galactic context.
    
    \item At redshift $z\lesssim 1$ we statistically match the distribution of impact parameters with metallicity based on the analytical F08+K17 model for DLAs. For higher redshifts, and when including the contribution from sub-DLAs, we find a greater statistical miss-match and excess detection probability beyond the envelope of the analytical model.
    
    \item To understand the physical origin of the detection-fractions, we perform a disc-halo separation using the baryonic angular momentum vector and the density profile of neutral gas. We find that $\sim 80\%$ of DLAs originate in extended \ion{H}{i} discs, with the remaining $20\%$ originating in the halo. Including sub-DLAs increases the relative contribution of the halo to $40\%$. We note that this separation relies on the existence of a clearly defined \ion{H}{i} disc. 
    
    \item We find a $3.2\sigma$ significance redshift evolution in the mean impact parameters of sub-DLA and DLA LOS. At high redshifts the mean impact parameters overlap, but separate at successively lower redshifts with sub-DLA LOS on average identified at a factor two larger projected separations than DLAs at $z\sim 0.4$. At all redshifts, our simulations can match the  observational data.
    
    \item We find a $3.5\sigma$ significance anti-correlation in differential covering-fraction of DLA to sub-DLA LOSs with stellar mass. This suggests an observational selection effect that causes the preferential detection of sub-DLAs in more massive galaxies in the low-redshift universe.

\end{itemize}

Despite the remarkable agreement with observations - which may suggest that damped \ion{H}{i} absorption predominantly form in typical gas-rich disc galaxies - we caution against over-interpreting the results. Whereas our simulation qualitatively capture and explain the current set of observations, the results should be anchored in the next generation of zoom-in simulations performed across a large grid to account for environmental dependence and sample the underlying dark matter halo distribution.

\section*{Acknowledgements}
This work was supported by grant ID DFF-4090-00079.
The cosmic Dawn center is funded by the DNRF. This work used the COSMA Data Centric system at Durham University, operated by the Institute for Computational Cosmology on behalf of the STFC DiRAC HPC Facility (www.dirac.ac.uk). This equipment was funded by a BIS National E-infrastructure capital grant ST/K00042X/1, DiRAC Operations grant ST/K003267/1 and Durham University. DiRAC is part of the National E-Infrastructure. OA acknowledges support from the Swedish Research Council (grant 2014- 5791). OA and FR acknowledge support from the Knut and Alice Wallenberg Foundation. 



\bibliographystyle{mnras}
\bibliography{refs.bib} 

\begin{thebibliography}{}
\makeatletter
\relax
\def\mn@urlcharsother{\let\do\@makeother \do\$\do\&\do\#\do\^\do\_\do\%\do\~}
\def\mn@doi{\begingroup\mn@urlcharsother \@ifnextchar [ {\mn@doi@}
  {\mn@doi@[]}}
\def\mn@doi@[#1]#2{\def\@tempa{#1}\ifx\@tempa\@empty \href
  {http://dx.doi.org/#2} {doi:#2}\else \href {http://dx.doi.org/#2} {#1}\fi
  \endgroup}
\def\mn@eprint#1#2{\mn@eprint@#1:#2::\@nil}
\def\mn@eprint@arXiv#1{\href {http://arxiv.org/abs/#1} {{\tt arXiv:#1}}}
\def\mn@eprint@dblp#1{\href {http://dblp.uni-trier.de/rec/bibtex/#1.xml}
  {dblp:#1}}
\def\mn@eprint@#1:#2:#3:#4\@nil{\def\@tempa {#1}\def\@tempb {#2}\def\@tempc
  {#3}\ifx \@tempc \@empty \let \@tempc \@tempb \let \@tempb \@tempa \fi \ifx
  \@tempb \@empty \def\@tempb {arXiv}\fi \@ifundefined
  {mn@eprint@\@tempb}{\@tempb:\@tempc}{\expandafter \expandafter \csname
  mn@eprint@\@tempb\endcsname \expandafter{\@tempc}}}

\bibitem[\protect\citeauthoryear{{Agertz} \& {Kravtsov}}{{Agertz} \&
  {Kravtsov}}{2015}]{AgertzKravtsov2015}
{Agertz} O.,  {Kravtsov} A.~V.,  2015, \mn@doi [\apj]
  {10.1088/0004-637X/804/1/18}, \href
  {http://adsabs.harvard.edu/abs/2015ApJ...804...18A} {804, 18}

\bibitem[\protect\citeauthoryear{{Agertz} \& {Kravtsov}}{{Agertz} \&
  {Kravtsov}}{2016}]{agertzkravtsov2016}
{Agertz} O.,  {Kravtsov} A.~V.,  2016, \mn@doi [\apj]
  {10.3847/0004-637X/824/2/79}, \href
  {http://adsabs.harvard.edu/abs/2016ApJ...824...79A} {824, 79}

\bibitem[\protect\citeauthoryear{{Agertz}, {Teyssier}  \& {Moore}}{{Agertz}
  et~al.}{2009}]{Agertz09b}
{Agertz} O.,  {Teyssier} R.,   {Moore} B.,  2009, \mn@doi [\mnras]
  {10.1111/j.1745-3933.2009.00685.x}, \href
  {http://adsabs.harvard.edu/abs/2009MNRAS.397L..64A} {397, L64}

\bibitem[\protect\citeauthoryear{{Agertz}, {Kravtsov}, {Leitner}  \&
  {Gnedin}}{{Agertz} et~al.}{2013}]{Agertz2013}
{Agertz} O.,  {Kravtsov} A.~V.,  {Leitner} S.~N.,   {Gnedin} N.~Y.,  2013,
  \mn@doi [\apj] {10.1088/0004-637X/770/1/25}, \href
  {http://adsabs.harvard.edu/abs/2013ApJ...770...25A} {770, 25}

\bibitem[\protect\citeauthoryear{{Altay}, {Theuns}, {Schaye}, {Crighton}  \&
  {Dalla Vecchia}}{{Altay} et~al.}{2011}]{Altay2011}
{Altay} G.,  {Theuns} T.,  {Schaye} J.,  {Crighton} N.~H.~M.,   {Dalla Vecchia}
  C.,  2011, \mn@doi [\apjl] {10.1088/2041-8205/737/2/L37}, \href
  {http://adsabs.harvard.edu/abs/2011ApJ...737L..37A} {737, L37}

\bibitem[\protect\citeauthoryear{{Altay}, {Theuns}, {Schaye}, {Booth}  \&
  {Dalla Vecchia}}{{Altay} et~al.}{2013}]{Altay2013}
{Altay} G.,  {Theuns} T.,  {Schaye} J.,  {Booth} C.~M.,   {Dalla Vecchia} C.,
  2013, \mn@doi [\mnras] {10.1093/mnras/stt1765}, \href
  {http://adsabs.harvard.edu/abs/2013MNRAS.436.2689A} {436, 2689}

\bibitem[\protect\citeauthoryear{{Asplund}, {Grevesse}, {Sauval}  \&
  {Scott}}{{Asplund} et~al.}{2009}]{Asplund2009}
{Asplund} M.,  {Grevesse} N.,  {Sauval} A.~J.,   {Scott} P.,  2009, \mn@doi
  [\araa] {10.1146/annurev.astro.46.060407.145222}, \href
  {http://adsabs.harvard.edu/abs/2009ARA%26A..47..481A} {47, 481}

\bibitem[\protect\citeauthoryear{{Aubert} \& {Teyssier}}{{Aubert} \&
  {Teyssier}}{2010}]{AubertTeyssier2010}
{Aubert} D.,  {Teyssier} R.,  2010, \mn@doi [\apj]
  {10.1088/0004-637X/724/1/244}, \href
  {http://adsabs.harvard.edu/abs/2010ApJ...724..244A} {724, 244}

\bibitem[\protect\citeauthoryear{{Augustin} et~al.,}{{Augustin}
  et~al.}{2018}]{Augustin2018}
{Augustin} R.,  et~al., 2018, \mn@doi [\mnras] {10.1093/mnras/sty1287}, \href
  {http://adsabs.harvard.edu/abs/2018MNRAS.478.3120A} {478, 3120}

\bibitem[\protect\citeauthoryear{{Bird}, {Vogelsberger}, {Haehnelt}, {Sijacki},
  {Genel}, {Torrey}, {Springel}  \& {Hernquist}}{{Bird}
  et~al.}{2014}]{Bird2014}
{Bird} S.,  {Vogelsberger} M.,  {Haehnelt} M.,  {Sijacki} D.,  {Genel} S.,
  {Torrey} P.,  {Springel} V.,   {Hernquist} L.,  2014, \mn@doi [\mnras]
  {10.1093/mnras/stu1923}, \href
  {http://adsabs.harvard.edu/abs/2014MNRAS.445.2313B} {445, 2313}

\bibitem[\protect\citeauthoryear{{Bird}, {Haehnelt}, {Neeleman}, {Genel},
  {Vogelsberger}  \& {Hernquist}}{{Bird} et~al.}{2015}]{Bird2015}
{Bird} S.,  {Haehnelt} M.,  {Neeleman} M.,  {Genel} S.,  {Vogelsberger} M.,
  {Hernquist} L.,  2015, \mn@doi [\mnras] {10.1093/mnras/stu2542}, \href
  {http://adsabs.harvard.edu/abs/2015MNRAS.447.1834B} {447, 1834}

\bibitem[\protect\citeauthoryear{{Blondin}, {Wright}, {Borkowski}  \&
  {Reynolds}}{{Blondin} et~al.}{1998}]{Blondin1998}
{Blondin} J.~M.,  {Wright} E.~B.,  {Borkowski} K.~J.,   {Reynolds} S.~P.,
  1998, \mn@doi [\apj] {10.1086/305708}, \href
  {http://adsabs.harvard.edu/abs/1998ApJ...500..342B} {500, 342}

\bibitem[\protect\citeauthoryear{{Chabrier}}{{Chabrier}}{2003}]{chabrier03}
{Chabrier} G.,  2003, \mn@doi [\pasp] {10.1086/376392}, \href
  {http://adsabs.harvard.edu/abs/2003PASP..115..763C} {115, 763}

\bibitem[\protect\citeauthoryear{{Christensen}, {Wisotzki}, {Roth},
  {S{\'a}nchez}, {Kelz}  \& {Jahnke}}{{Christensen}
  et~al.}{2007}]{Christensen2007}
{Christensen} L.,  {Wisotzki} L.,  {Roth} M.~M.,  {S{\'a}nchez} S.~F.,  {Kelz}
  A.,   {Jahnke} K.,  2007, \mn@doi [\aap] {10.1051/0004-6361:20066410}, \href
  {http://adsabs.harvard.edu/abs/2007A%26A...468..587C} {468, 587}

\bibitem[\protect\citeauthoryear{{Christensen}, {M\o ller}, {Rhodin}, {Heintz}
  \& {Fynbo}}{{Christensen} et~al.}{2019}]{Christensen2019}
{Christensen} L.,  {M\o ller} P.,  {Rhodin} N. H.~P.,  {Heintz} K.,   {Fynbo}
  J. P.~U.,  2019, \mnras

\bibitem[\protect\citeauthoryear{{Cioffi}, {McKee}  \& {Bertschinger}}{{Cioffi}
  et~al.}{1988}]{Cioffi1988}
{Cioffi} D.~F.,  {McKee} C.~F.,   {Bertschinger} E.,  1988, \mn@doi [\apj]
  {10.1086/166834}, \href {http://adsabs.harvard.edu/abs/1988ApJ...334..252C}
  {334, 252}

\bibitem[\protect\citeauthoryear{{Dessauges-Zavadsky}, {Ellison}  \&
  {Murphy}}{{Dessauges-Zavadsky} et~al.}{2009}]{Dessauges-Zavadsky2009}
{Dessauges-Zavadsky} M.,  {Ellison} S.~L.,   {Murphy} M.~T.,  2009, \mn@doi
  [\mnras] {10.1111/j.1745-3933.2009.00662.x}, \href
  {http://adsabs.harvard.edu/abs/2009MNRAS.396L..61D} {396, L61}

\bibitem[\protect\citeauthoryear{{Eisenstein} \& {Hut}}{{Eisenstein} \&
  {Hut}}{1998}]{HOP1998}
{Eisenstein} D.~J.,  {Hut} P.,  1998, \mn@doi [\apj] {10.1086/305535}, \href
  {http://adsabs.harvard.edu/abs/1998ApJ...498..137E} {498, 137}

\bibitem[\protect\citeauthoryear{{Erkal}, {Gnedin}  \& {Kravtsov}}{{Erkal}
  et~al.}{2012}]{Erkal2012}
{Erkal} D.,  {Gnedin} N.~Y.,   {Kravtsov} A.~V.,  2012, \mn@doi [\apj]
  {10.1088/0004-637X/761/1/54}, \href
  {http://adsabs.harvard.edu/abs/2012ApJ...761...54E} {761, 54}

\bibitem[\protect\citeauthoryear{{Faucher-Gigu{\`e}re}, {Hopkins}, {Kere{\v
  s}}, {Muratov}, {Quataert}  \& {Murray}}{{Faucher-Gigu{\`e}re}
  et~al.}{2015}]{FaucherGiguere2015}
{Faucher-Gigu{\`e}re} C.-A.,  {Hopkins} P.~F.,  {Kere{\v s}} D.,  {Muratov}
  A.~L.,  {Quataert} E.,   {Murray} N.,  2015, \mn@doi [\mnras]
  {10.1093/mnras/stv336}, \href
  {http://adsabs.harvard.edu/abs/2015MNRAS.449..987F} {449, 987}

\bibitem[\protect\citeauthoryear{{Freudling}, {M\o ller}, {Krogager}  \&
  {Christensen}}{{Freudling} et~al.}{2019}]{Freudling2019}
{Freudling} W.,  {M\o ller} P.,  {Krogager} J.-K.,   {Christensen} L.,  2019

\bibitem[\protect\citeauthoryear{{Fynbo}, {Prochaska}, {Sommer-Larsen},
  {Dessauges-Zavadsky}  \& {M{\o}ller}}{{Fynbo} et~al.}{2008}]{Fynbo2008}
{Fynbo} J.~P.~U.,  {Prochaska} J.~X.,  {Sommer-Larsen} J.,
  {Dessauges-Zavadsky} M.,   {M{\o}ller} P.,  2008, \mn@doi [\apj]
  {10.1086/589555}, \href {http://adsabs.harvard.edu/abs/2008ApJ...683..321F}
  {683, 321}

\bibitem[\protect\citeauthoryear{{Fynbo} et~al.,}{{Fynbo}
  et~al.}{2010}]{Fynbo2010}
{Fynbo} J.~P.~U.,  et~al., 2010, \mn@doi [\mnras]
  {10.1111/j.1365-2966.2010.17294.x}, \href
  {http://adsabs.harvard.edu/abs/2010MNRAS.408.2128F} {408, 2128}

\bibitem[\protect\citeauthoryear{{Fynbo} et~al.,}{{Fynbo}
  et~al.}{2011}]{Fynbo2011}
{Fynbo} J.~P.~U.,  et~al., 2011, \mn@doi [\mnras]
  {10.1111/j.1365-2966.2011.18318.x}, \href
  {http://adsabs.harvard.edu/abs/2011MNRAS.413.2481F} {413, 2481}

\bibitem[\protect\citeauthoryear{{Fynbo}, {Krogager}, {Venemans}, {Noterdaeme},
  {Vestergaard}, {M{\o}ller}, {Ledoux}  \& {Geier}}{{Fynbo}
  et~al.}{2013a}]{Fynbo2013b}
{Fynbo} J.~P.~U.,  {Krogager} J.-K.,  {Venemans} B.,  {Noterdaeme} P.,
  {Vestergaard} M.,  {M{\o}ller} P.,  {Ledoux} C.,   {Geier} S.,  2013a,
  \mn@doi [\apjs] {10.1088/0067-0049/204/1/6}, \href
  {http://adsabs.harvard.edu/abs/2013ApJS..204....6F} {204, 6}

\bibitem[\protect\citeauthoryear{{Fynbo} et~al.,}{{Fynbo}
  et~al.}{2013b}]{Fynbo2013}
{Fynbo} J.~P.~U.,  et~al., 2013b, \mn@doi [\mnras] {10.1093/mnras/stt1579},
  \href {http://adsabs.harvard.edu/abs/2013MNRAS.436..361F} {436, 361}

\bibitem[\protect\citeauthoryear{{Fynbo} et~al.,}{{Fynbo}
  et~al.}{2017}]{Fynbo2017}
{Fynbo} J.~P.~U.,  et~al., 2017, \mn@doi [\aap] {10.1051/0004-6361/201730726},
  \href {http://adsabs.harvard.edu/abs/2017A%26A...606A..13F} {606, A13}

\bibitem[\protect\citeauthoryear{{Haardt} \& {Madau}}{{Haardt} \&
  {Madau}}{1996}]{haardtmadau96}
{Haardt} F.,  {Madau} P.,  1996, \mn@doi [\apj] {10.1086/177035}, \href
  {http://adsabs.harvard.edu/abs/1996ApJ...461...20H} {461, 20}

\bibitem[\protect\citeauthoryear{Haehnelt, {Steinmetz}  \& {Rauch}}{Haehnelt
  et~al.}{1998}]{Haehnelt1998}
Haehnelt M.~G.,  {Steinmetz} M.,   {Rauch} M.,  1998, \mn@doi [\apj]
  {10.1086/305323}, \href {http://adsabs.harvard.edu/abs/1998ApJ...495..647H}
  {495, 647}

\bibitem[\protect\citeauthoryear{{Hahn} \& {Abel}}{{Hahn} \&
  {Abel}}{2011}]{music2011}
{Hahn} O.,  {Abel} T.,  2011, \mn@doi [\mnras]
  {10.1111/j.1365-2966.2011.18820.x}, \href
  {http://adsabs.harvard.edu/abs/2011MNRAS.415.2101H} {415, 2101}

\bibitem[\protect\citeauthoryear{{Hopkins}, {Kere{\v s}}, {O{\~n}orbe},
  {Faucher-Gigu{\`e}re}, {Quataert}, {Murray}  \& {Bullock}}{{Hopkins}
  et~al.}{2014}]{Hopkins2014}
{Hopkins} P.~F.,  {Kere{\v s}} D.,  {O{\~n}orbe} J.,  {Faucher-Gigu{\`e}re}
  C.-A.,  {Quataert} E.,  {Murray} N.,   {Bullock} J.~S.,  2014, \mn@doi
  [\mnras] {10.1093/mnras/stu1738}, \href
  {http://adsabs.harvard.edu/abs/2014MNRAS.445..581H} {445, 581}

\bibitem[\protect\citeauthoryear{{Hummels} et~al.,}{{Hummels}
  et~al.}{2018}]{Hummels2018}
{Hummels} C.~B.,  et~al., 2018, arXiv e-prints, \href
  {http://adsabs.harvard.edu/abs/2018arXiv181112410H} {}

\bibitem[\protect\citeauthoryear{{Khare}, {Kulkarni}, {P{\'e}roux}, {York},
  {Lauroesch}  \& {Meiring}}{{Khare} et~al.}{2007}]{Khare2007}
{Khare} P.,  {Kulkarni} V.~P.,  {P{\'e}roux} C.,  {York} D.~G.,  {Lauroesch}
  J.~T.,   {Meiring} J.~D.,  2007, \mn@doi [\aap] {10.1051/0004-6361:20066186},
  \href {http://adsabs.harvard.edu/abs/2007A%26A...464..487K} {464, 487}

\bibitem[\protect\citeauthoryear{{Kim} \& {Ostriker}}{{Kim} \&
  {Ostriker}}{2015}]{KimOstriker2015}
{Kim} C.-G.,  {Ostriker} E.~C.,  2015, \mn@doi [\apj]
  {10.1088/0004-637X/802/2/99}, \href
  {http://adsabs.harvard.edu/abs/2015ApJ...802...99K} {802, 99}

\bibitem[\protect\citeauthoryear{{Kim}, {Abel}  \& {Agertz}}{{Kim}
  et~al.}{2014}]{agora}
{Kim} J.-h.,  {Abel} T.,   {Agertz} O. e.~a.,  2014, \mn@doi [\apjs]
  {10.1088/0067-0049/210/1/14}, \href
  {http://adsabs.harvard.edu/abs/2014ApJS..210...14K} {210, 14}

\bibitem[\protect\citeauthoryear{{Kim} et~al.,}{{Kim} et~al.}{2016}]{agora2}
{Kim} J.-h.,  et~al., 2016, \mn@doi [\apj] {10.3847/1538-4357/833/2/202}, \href
  {http://adsabs.harvard.edu/abs/2016ApJ...833..202K} {833, 202}

\bibitem[\protect\citeauthoryear{{Krogager}, {Fynbo}, {M{\o}ller}, {Ledoux},
  {Noterdaeme}, {Christensen}, {Milvang-Jensen}  \& {Sparre}}{{Krogager}
  et~al.}{2012}]{Krogager2012}
{Krogager} J.-K.,  {Fynbo} J.~P.~U.,  {M{\o}ller} P.,  {Ledoux} C.,
  {Noterdaeme} P.,  {Christensen} L.,  {Milvang-Jensen} B.,   {Sparre} M.,
  2012, \mn@doi [\mnras] {10.1111/j.1745-3933.2012.01272.x}, \href
  {http://adsabs.harvard.edu/abs/2012MNRAS.424L...1K} {424, L1}

\bibitem[\protect\citeauthoryear{{Krogager} et~al.,}{{Krogager}
  et~al.}{2013}]{Krogager2013}
{Krogager} J.-K.,  et~al., 2013, \mn@doi [\mnras] {10.1093/mnras/stt955}, \href
  {http://adsabs.harvard.edu/abs/2013MNRAS.433.3091K} {433, 3091}

\bibitem[\protect\citeauthoryear{{Krogager} et~al.,}{{Krogager}
  et~al.}{2015}]{Krogager2015}
{Krogager} J.-K.,  et~al., 2015, \mn@doi [\apjs] {10.1088/0067-0049/217/1/5},
  \href {http://adsabs.harvard.edu/abs/2015ApJS..217....5K} {217, 5}

\bibitem[\protect\citeauthoryear{{Krogager} et~al.,}{{Krogager}
  et~al.}{2016}]{Krogager2016}
{Krogager} J.-K.,  et~al., 2016, \mn@doi [\apj] {10.3847/0004-637X/832/1/49},
  \href {http://adsabs.harvard.edu/abs/2016ApJ...832...49K} {832, 49}

\bibitem[\protect\citeauthoryear{{Krogager}, {M{\o}ller}, {Fynbo}  \&
  {Noterdaeme}}{{Krogager} et~al.}{2017}]{Krogager2017}
{Krogager} J.-K.,  {M{\o}ller} P.,  {Fynbo} J.~P.~U.,   {Noterdaeme} P.,  2017,
  \mn@doi [\mnras] {10.1093/mnras/stx1011}, \href
  {http://adsabs.harvard.edu/abs/2017MNRAS.469.2959K} {469, 2959}

\bibitem[\protect\citeauthoryear{Krumholz, McKee  \& Tumlinson}{Krumholz
  et~al.}{2008}]{Krumholz2008}
Krumholz M.~R.,  McKee C.~F.,   Tumlinson J.,  2008, Astrophyiscal Journal,
  689, 865

\bibitem[\protect\citeauthoryear{Krumholz, McKee  \& Tumlinson}{Krumholz
  et~al.}{2009}]{Krumholz09}
Krumholz M.~R.,  McKee C.~F.,   Tumlinson J.,  2009, Astrophyiscal Journal,
  693, 216

\bibitem[\protect\citeauthoryear{{Kulkarni}, {Khare}, {Som}, {Meiring}, {York},
  {P{\'e}roux}  \& {Lauroesch}}{{Kulkarni} et~al.}{2010}]{Kulkarni2010}
{Kulkarni} V.~P.,  {Khare} P.,  {Som} D.,  {Meiring} J.,  {York} D.~G.,
  {P{\'e}roux} C.,   {Lauroesch} J.~T.,  2010, \mn@doi [\na]
  {10.1016/j.newast.2010.05.006}, \href
  {http://adsabs.harvard.edu/abs/2010NewA...15..735K} {15, 735}

\bibitem[\protect\citeauthoryear{{Ledoux}, {Petitjean}, {Fynbo}, {M{\o}ller}
  \& {Srianand}}{{Ledoux} et~al.}{2006}]{Ledoux2006}
{Ledoux} C.,  {Petitjean} P.,  {Fynbo} J.~P.~U.,  {M{\o}ller} P.,   {Srianand}
  R.,  2006, \mn@doi [\aap] {10.1051/0004-6361:20054242}, \href
  {http://adsabs.harvard.edu/abs/2006A%26A...457...71L} {457, 71}

\bibitem[\protect\citeauthoryear{{Leitherer} et~al.,}{{Leitherer}
  et~al.}{1999}]{Leitherer1999}
{Leitherer} C.,  et~al., 1999, \mn@doi [\apjs] {10.1086/313233}, \href
  {http://adsabs.harvard.edu/abs/1999ApJS..123....3L} {123, 3}

\bibitem[\protect\citeauthoryear{{Liang}, {Kravtsov}  \& {Agertz}}{{Liang}
  et~al.}{2015}]{Liang2015}
{Liang} C.~J.,  {Kravtsov} A.~V.,   {Agertz} O.,  2015, {\mnras} submitted
  (arxiv/1507.07002), \href {http://adsabs.harvard.edu/abs/2015arXiv150707002L}
  {}

\bibitem[\protect\citeauthoryear{{Martizzi}, {Faucher-Gigu{\`e}re}  \&
  {Quataert}}{{Martizzi} et~al.}{2015}]{Martizzi2015}
{Martizzi} D.,  {Faucher-Gigu{\`e}re} C.-A.,   {Quataert} E.,  2015, \mn@doi
  [\mnras] {10.1093/mnras/stv562}, \href
  {http://adsabs.harvard.edu/abs/2015MNRAS.450..504M} {450, 504}

\bibitem[\protect\citeauthoryear{{McCourt}, {Oh}, {O'Leary}  \&
  {Madigan}}{{McCourt} et~al.}{2018}]{McCourt2018}
{McCourt} M.,  {Oh} S.~P.,  {O'Leary} R.,   {Madigan} A.-M.,  2018, \mn@doi
  [\mnras] {10.1093/mnras/stx2687}, \href
  {http://adsabs.harvard.edu/abs/2018MNRAS.473.5407M} {473, 5407}

\bibitem[\protect\citeauthoryear{{Meiring}, {Lauroesch}, {Haberzettl},
  {Kulkarni}, {P{\'e}roux}, {Khare}  \& {York}}{{Meiring}
  et~al.}{2011}]{Meiring2011}
{Meiring} J.~D.,  {Lauroesch} J.~T.,  {Haberzettl} L.,  {Kulkarni} V.~P.,
  {P{\'e}roux} C.,  {Khare} P.,   {York} D.~G.,  2011, \mn@doi [\mnras]
  {10.1111/j.1365-2966.2010.17625.x}, \href
  {http://adsabs.harvard.edu/abs/2011MNRAS.410.2516M} {410, 2516}

\bibitem[\protect\citeauthoryear{{Mo}, {Mao}  \& {White}}{{Mo}
  et~al.}{1998}]{Mo1998}
{Mo} H.~J.,  {Mao} S.,   {White} S.~D.~M.,  1998, \mn@doi [\mnras]
  {10.1046/j.1365-8711.1998.01227.x}, \href
  {http://adsabs.harvard.edu/abs/1998MNRAS.295..319M} {295, 319}

\bibitem[\protect\citeauthoryear{{M\o ller} \& {Christensen}}{{M\o ller} \&
  {Christensen}}{2019}]{Moller2019}
{M\o ller} P.,  {Christensen} L.,  2019, \mnras

\bibitem[\protect\citeauthoryear{{M{\o}ller} \& {Warren}}{{M{\o}ller} \&
  {Warren}}{1998}]{Moller1998}
{M{\o}ller} P.,  {Warren} S.~J.,  1998, \mn@doi [\mnras]
  {10.1046/j.1365-8711.1998.01749.x}, \href
  {http://adsabs.harvard.edu/abs/1998MNRAS.299..661M} {299, 661}

\bibitem[\protect\citeauthoryear{{M{\o}ller}, {Fynbo}, {Ledoux}  \&
  {Nilsson}}{{M{\o}ller} et~al.}{2013}]{Moller2013}
{M{\o}ller} P.,  {Fynbo} J.~P.~U.,  {Ledoux} C.,   {Nilsson} K.~K.,  2013,
  \mn@doi [\mnras] {10.1093/mnras/stt067}, \href
  {http://adsabs.harvard.edu/abs/2013MNRAS.430.2680M} {430, 2680}

\bibitem[\protect\citeauthoryear{{Monier}, {Turnshek}  \& {Rao}}{{Monier}
  et~al.}{2009}]{Monier2009}
{Monier} E.~M.,  {Turnshek} D.~A.,   {Rao} S.,  2009, \mn@doi [\mnras]
  {10.1111/j.1365-2966.2009.15000.x}, \href
  {http://adsabs.harvard.edu/abs/2009MNRAS.397..943M} {397, 943}

\bibitem[\protect\citeauthoryear{{Neeleman}, {Wolfe}, {Prochaska}  \&
  {Rafelski}}{{Neeleman} et~al.}{2013}]{Neeleman2013}
{Neeleman} M.,  {Wolfe} A.~M.,  {Prochaska} J.~X.,   {Rafelski} M.,  2013,
  \mn@doi [\apj] {10.1088/0004-637X/769/1/54}, \href
  {http://adsabs.harvard.edu/abs/2013ApJ...769...54N} {769, 54}

\bibitem[\protect\citeauthoryear{{Noterdaeme}, {Petitjean}, {Ledoux}  \&
  {Srianand}}{{Noterdaeme} et~al.}{2009}]{Noterdaeme2009}
{Noterdaeme} P.,  {Petitjean} P.,  {Ledoux} C.,   {Srianand} R.,  2009, \mn@doi
  [\aap] {10.1051/0004-6361/200912768}, \href
  {http://adsabs.harvard.edu/abs/2009A%26A...505.1087N} {505, 1087}

\bibitem[\protect\citeauthoryear{{Noterdaeme} et~al.,}{{Noterdaeme}
  et~al.}{2012}]{Noterdaeme2012b}
{Noterdaeme} P.,  et~al., 2012, \mn@doi [\aap] {10.1051/0004-6361/201220259},
  \href {http://adsabs.harvard.edu/abs/2012A%26A...547L...1N} {547, L1}

\bibitem[\protect\citeauthoryear{{Noterdaeme}, {Petitjean}  \&
  {Srianand}}{{Noterdaeme} et~al.}{2015}]{Noterdaeme2015}
{Noterdaeme} P.,  {Petitjean} P.,   {Srianand} R.,  2015, \mn@doi [\aap]
  {10.1051/0004-6361/201526018}, \href
  {http://adsabs.harvard.edu/abs/2015A%26A...578L...5N} {578, L5}

\bibitem[\protect\citeauthoryear{{Padoan}, {Haugb{\o}lle}  \&
  {Nordlund}}{{Padoan} et~al.}{2012}]{Padoan2012}
{Padoan} P.,  {Haugb{\o}lle} T.,   {Nordlund} {\AA}.,  2012, \mn@doi [\apjl]
  {10.1088/2041-8205/759/2/L27}, \href
  {http://adsabs.harvard.edu/abs/2012ApJ...759L..27P} {759, L27}

\bibitem[\protect\citeauthoryear{{P{\'e}roux}, {Dessauges-Zavadsky},
  {D'Odorico}, {Kim}  \& {McMahon}}{{P{\'e}roux} et~al.}{2003}]{Peroux2003}
{P{\'e}roux} C.,  {Dessauges-Zavadsky} M.,  {D'Odorico} S.,  {Kim} T.-S.,
  {McMahon} R.~G.,  2003, \mn@doi [\mnras] {10.1046/j.1365-8711.2003.06952.x},
  \href {http://adsabs.harvard.edu/abs/2003MNRAS.345..480P} {345, 480}

\bibitem[\protect\citeauthoryear{{P{\'e}roux}, {Bouch{\'e}}, {Kulkarni}, {York}
   \& {Vladilo}}{{P{\'e}roux} et~al.}{2012}]{Peroux2012}
{P{\'e}roux} C.,  {Bouch{\'e}} N.,  {Kulkarni} V.~P.,  {York} D.~G.,
  {Vladilo} G.,  2012, \mn@doi [\mnras] {10.1111/j.1365-2966.2011.19947.x},
  \href {http://adsabs.harvard.edu/abs/2012MNRAS.419.3060P} {419, 3060}

\bibitem[\protect\citeauthoryear{{Pettini}, {Shapley}, {Steidel}, {Cuby},
  {Dickinson}, {Moorwood}, {Adelberger}  \& {Giavalisco}}{{Pettini}
  et~al.}{2001}]{Pettini2001}
{Pettini} M.,  {Shapley} A.~E.,  {Steidel} C.~C.,  {Cuby} J.-G.,  {Dickinson}
  M.,  {Moorwood} A.~F.~M.,  {Adelberger} K.~L.,   {Giavalisco} M.,  2001,
  \mn@doi [\apj] {10.1086/321403}, \href
  {http://adsabs.harvard.edu/abs/2001ApJ...554..981P} {554, 981}

\bibitem[\protect\citeauthoryear{{Pontzen} et~al.,}{{Pontzen}
  et~al.}{2008}]{Pontzen2008}
{Pontzen} A.,  et~al., 2008, \mn@doi [\mnras]
  {10.1111/j.1365-2966.2008.13782.x}, \href
  {http://adsabs.harvard.edu/abs/2008MNRAS.390.1349P} {390, 1349}

\bibitem[\protect\citeauthoryear{{Power}, {Navarro}, {Jenkins}, {Frenk},
  {White}, {Springel}, {Stadel}  \& {Quinn}}{{Power} et~al.}{2003}]{Power2003}
{Power} C.,  {Navarro} J.~F.,  {Jenkins} A.,  {Frenk} C.~S.,  {White} S.~D.~M.,
   {Springel} V.,  {Stadel} J.,   {Quinn} T.,  2003, \mn@doi [\mnras]
  {10.1046/j.1365-8711.2003.05925.x}, \href
  {http://adsabs.harvard.edu/abs/2003MNRAS.338...14P} {338, 14}

\bibitem[\protect\citeauthoryear{{Prochaska} \& {Wolfe}}{{Prochaska} \&
  {Wolfe}}{1997}]{Prochaska1997}
{Prochaska} J.~X.,  {Wolfe} A.~M.,  1997, \mn@doi [\apj] {10.1086/304591},
  \href {http://adsabs.harvard.edu/abs/1997ApJ...487...73P} {487, 73}

\bibitem[\protect\citeauthoryear{{Prochaska}, {Gawiser}, {Wolfe}, {Castro}  \&
  {Djorgovski}}{{Prochaska} et~al.}{2003}]{Prochaska2003}
{Prochaska} J.~X.,  {Gawiser} E.,  {Wolfe} A.~M.,  {Castro} S.,   {Djorgovski}
  S.~G.,  2003, \mn@doi [\apjl] {10.1086/378945}, \href
  {http://adsabs.harvard.edu/abs/2003ApJ...595L...9P} {595, L9}

\bibitem[\protect\citeauthoryear{{Prochaska}, {Madau}, {O'Meara}  \&
  {Fumagalli}}{{Prochaska} et~al.}{2014}]{Prochaska2014}
{Prochaska} J.~X.,  {Madau} P.,  {O'Meara} J.~M.,   {Fumagalli} M.,  2014,
  \mn@doi [\mnras] {10.1093/mnras/stt2218}, \href
  {http://adsabs.harvard.edu/abs/2014MNRAS.438..476P} {438, 476}

\bibitem[\protect\citeauthoryear{{Rafelski}, {Neeleman}, {Fumagalli}, {Wolfe}
  \& {Prochaska}}{{Rafelski} et~al.}{2014}]{Rafelski2014}
{Rafelski} M.,  {Neeleman} M.,  {Fumagalli} M.,  {Wolfe} A.~M.,   {Prochaska}
  J.~X.,  2014, \mn@doi [\apjl] {10.1088/2041-8205/782/2/L29}, \href
  {http://adsabs.harvard.edu/abs/2014ApJ...782L..29R} {782, L29}

\bibitem[\protect\citeauthoryear{{Rahmani} et~al.,}{{Rahmani}
  et~al.}{2016}]{Rahmani2016}
{Rahmani} H.,  et~al., 2016, \mn@doi [\mnras] {10.1093/mnras/stw1965}, \href
  {http://adsabs.harvard.edu/abs/2016MNRAS.463..980R} {463, 980}

\bibitem[\protect\citeauthoryear{{Rahmati} \& {Schaye}}{{Rahmati} \&
  {Schaye}}{2014}]{Rahmati2014}
{Rahmati} A.,  {Schaye} J.,  2014, \mn@doi [\mnras] {10.1093/mnras/stt2235},
  \href {http://adsabs.harvard.edu/abs/2014MNRAS.438..529R} {438, 529}

\bibitem[\protect\citeauthoryear{Raiteri, Villata  \& Navarro}{Raiteri
  et~al.}{1996}]{Raiteri1996}
Raiteri C.~M.,  Villata M.,   Navarro J.~F.,  1996, A{\&}A, 315, 105

\bibitem[\protect\citeauthoryear{{Rao}, {Belfort-Mihalyi}, {Turnshek},
  {Monier}, {Nestor}  \& {Quider}}{{Rao} et~al.}{2011}]{Rao2011}
{Rao} S.~M.,  {Belfort-Mihalyi} M.,  {Turnshek} D.~A.,  {Monier} E.~M.,
  {Nestor} D.~B.,   {Quider} A.,  2011, \mn@doi [\mnras]
  {10.1111/j.1365-2966.2011.19119.x}, \href
  {http://adsabs.harvard.edu/abs/2011MNRAS.416.1215R} {416, 1215}

\bibitem[\protect\citeauthoryear{{Reddy} \& {Steidel}}{{Reddy} \&
  {Steidel}}{2009}]{Reddy2009}
{Reddy} N.~A.,  {Steidel} C.~C.,  2009, \mn@doi [\apj]
  {10.1088/0004-637X/692/1/778}, \href
  {http://adsabs.harvard.edu/abs/2009ApJ...692..778R} {692, 778}

\bibitem[\protect\citeauthoryear{{Reddy}, {Steidel}, {Pettini}, {Adelberger},
  {Shapley}, {Erb}  \& {Dickinson}}{{Reddy} et~al.}{2008}]{Reddy2008}
{Reddy} N.~A.,  {Steidel} C.~C.,  {Pettini} M.,  {Adelberger} K.~L.,  {Shapley}
  A.~E.,  {Erb} D.~K.,   {Dickinson} M.,  2008, \mn@doi [\apjs]
  {10.1086/521105}, \href {http://adsabs.harvard.edu/abs/2008ApJS..175...48R}
  {175, 48}

\bibitem[\protect\citeauthoryear{{Rhodin}, {Christensen}, {M{\o}ller}, {Zafar}
  \& {Fynbo}}{{Rhodin} et~al.}{2018}]{Rhodin2018}
{Rhodin} N.~H.~P.,  {Christensen} L.,  {M{\o}ller} P.,  {Zafar} T.,   {Fynbo}
  J.~P.~U.,  2018, preprint, \href
  {http://adsabs.harvard.edu/abs/2018arXiv180701755R} {} (\mn@eprint {arXiv}
  {1807.01755})

\bibitem[\protect\citeauthoryear{{Romeo}, {Agertz}, {Moore}  \&
  {Stadel}}{{Romeo} et~al.}{2008}]{Romeo08}
{Romeo} A.~B.,  {Agertz} O.,  {Moore} B.,   {Stadel} J.,  2008, \mn@doi [\apj]
  {10.1086/591236}, \href {http://adsabs.harvard.edu/abs/2008ApJ...686....1R}
  {686, 1}

\bibitem[\protect\citeauthoryear{{Rosen} \& {Bregman}}{{Rosen} \&
  {Bregman}}{1995}]{rosenbregman95}
{Rosen} A.,  {Bregman} J.~N.,  1995, \mn@doi [\apj] {10.1086/175303}, \href
  {http://adsabs.harvard.edu/abs/1995ApJ...440..634R} {440, 634}

\bibitem[\protect\citeauthoryear{{Schaye} et~al.,}{{Schaye}
  et~al.}{2015}]{Schaye2015}
{Schaye} J.,  et~al., 2015, \mn@doi [\mnras] {10.1093/mnras/stu2058}, \href
  {http://adsabs.harvard.edu/abs/2015MNRAS.446..521S} {446, 521}

\bibitem[\protect\citeauthoryear{{Som}, {Kulkarni}, {Meiring}, {York},
  {P{\'e}roux}, {Lauroesch}, {Aller}  \& {Khare}}{{Som} et~al.}{2015}]{Som2015}
{Som} D.,  {Kulkarni} V.~P.,  {Meiring} J.,  {York} D.~G.,  {P{\'e}roux} C.,
  {Lauroesch} J.~T.,  {Aller} M.~C.,   {Khare} P.,  2015, \mn@doi [\apj]
  {10.1088/0004-637X/806/1/25}, \href
  {http://adsabs.harvard.edu/abs/2015ApJ...806...25S} {806, 25}

\bibitem[\protect\citeauthoryear{{Sutherland} \& {Dopita}}{{Sutherland} \&
  {Dopita}}{1993}]{sutherlanddopita93}
{Sutherland} R.~S.,  {Dopita} M.~A.,  1993, \mn@doi [\apjs] {10.1086/191823},
  \href {http://adsabs.harvard.edu/abs/1993ApJS...88..253S} {88, 253}

\bibitem[\protect\citeauthoryear{{Teyssier}}{{Teyssier}}{2002}]{teyssier02}
{Teyssier} R.,  2002, \mn@doi [\aap] {10.1051/0004-6361:20011817}, \href
  {http://adsabs.harvard.edu/abs/2002A%26A...385..337T} {385, 337}

\bibitem[\protect\citeauthoryear{{Thornton}, {Gaudlitz}, {Janka}  \&
  {Steinmetz}}{{Thornton} et~al.}{1998}]{Thornton1998}
{Thornton} K.,  {Gaudlitz} M.,  {Janka} H.-T.,   {Steinmetz} M.,  1998, \mn@doi
  [\apj] {10.1086/305704}, \href
  {http://adsabs.harvard.edu/abs/1998ApJ...500...95T} {500, 95}

\bibitem[\protect\citeauthoryear{{Vladilo} \& {P{\'e}roux}}{{Vladilo} \&
  {P{\'e}roux}}{2005}]{Vladilo2005}
{Vladilo} G.,  {P{\'e}roux} C.,  2005, \mn@doi [\aap]
  {10.1051/0004-6361:20041570}, \href
  {http://adsabs.harvard.edu/abs/2005A%26A...444..461V} {444, 461}

\bibitem[\protect\citeauthoryear{{Vogelsberger} et~al.,}{{Vogelsberger}
  et~al.}{2014}]{Vogelsberger2014}
{Vogelsberger} M.,  et~al., 2014, \mn@doi [\mnras] {10.1093/mnras/stu1536},
  \href {http://adsabs.harvard.edu/abs/2014MNRAS.444.1518V} {444, 1518}

\bibitem[\protect\citeauthoryear{{Wetzel}, {Hopkins}, {Kim},
  {Faucher-Gigu{\`e}re}, {Kere{\v s}}  \& {Quataert}}{{Wetzel}
  et~al.}{2016}]{Wetzel2016}
{Wetzel} A.~R.,  {Hopkins} P.~F.,  {Kim} J.-h.,  {Faucher-Gigu{\`e}re} C.-A.,
  {Kere{\v s}} D.,   {Quataert} E.,  2016, \mn@doi [\apjl]
  {10.3847/2041-8205/827/2/L23}, \href
  {http://adsabs.harvard.edu/abs/2016ApJ...827L..23W} {827, L23}

\bibitem[\protect\citeauthoryear{{Wise}, {Abel}, {Turk}, {Norman}  \&
  {Smith}}{{Wise} et~al.}{2012}]{Wise2012}
{Wise} J.~H.,  {Abel} T.,  {Turk} M.~J.,  {Norman} M.~L.,   {Smith} B.~D.,
  2012, \mn@doi [\mnras] {10.1111/j.1365-2966.2012.21809.x}, \href
  {http://adsabs.harvard.edu/abs/2012MNRAS.427..311W} {427, 311}

\bibitem[\protect\citeauthoryear{{Wolfe}, {Turnshek}, {Smith}  \&
  {Cohen}}{{Wolfe} et~al.}{1986}]{Wolfe1986}
{Wolfe} A.~M.,  {Turnshek} D.~A.,  {Smith} H.~E.,   {Cohen} R.~D.,  1986,
  \mn@doi [\apjs] {10.1086/191114}, \href
  {http://adsabs.harvard.edu/abs/1986ApJS...61..249W} {61, 249}

\bibitem[\protect\citeauthoryear{{Zafar} \& {Watson}}{{Zafar} \&
  {Watson}}{2013}]{Zafar2013b}
{Zafar} T.,  {Watson} D.,  2013, \mn@doi [\aap] {10.1051/0004-6361/201321413},
  \href {http://adsabs.harvard.edu/abs/2013A%26A...560A..26Z} {560, A26}

\bibitem[\protect\citeauthoryear{{Zafar}, {Popping}  \& {P{\'e}roux}}{{Zafar}
  et~al.}{2013}]{Zafar2013}
{Zafar} T.,  {Popping} A.,   {P{\'e}roux} C.,  2013, \mn@doi [\aap]
  {10.1051/0004-6361/201321153}, \href
  {http://adsabs.harvard.edu/abs/2013A%26A...556A.140Z} {556, A140}

\bibitem[\protect\citeauthoryear{{Zafar} et~al.,}{{Zafar}
  et~al.}{2015}]{Zafar2015}
{Zafar} T.,  et~al., 2015, \mn@doi [\aap] {10.1051/0004-6361/201526570}, \href
  {http://adsabs.harvard.edu/abs/2015A%26A...584A.100Z} {584, A100}

\bibitem[\protect\citeauthoryear{{Zwaan}, {van der Hulst}, {Briggs},
  {Verheijen}  \& {Ryan-Weber}}{{Zwaan} et~al.}{2005}]{Zwaan2005}
{Zwaan} M.~A.,  {van der Hulst} J.~M.,  {Briggs} F.~H.,  {Verheijen} M.~A.~W.,
   {Ryan-Weber} E.~V.,  2005, \mn@doi [\mnras]
  {10.1111/j.1365-2966.2005.09698.x}, \href
  {http://adsabs.harvard.edu/abs/2005MNRAS.364.1467Z} {364, 1467}

\makeatother
\end{thebibliography}






\bsp	
\label{lastpage}
\end{document}